\documentclass[11pt]{article}
\usepackage{amsfonts,amsmath,amssymb,amsthm}
\usepackage[pdftex]{graphicx}
\usepackage[hmargin=1in,vmargin=1in]{geometry}
\usepackage{natbib}
\usepackage{enumerate}
\usepackage{caption}
\usepackage[usenames,dvipsnames]{xcolor}
\usepackage{footmisc,fancyvrb} % enabling footnotes.
\usepackage[colorlinks,citecolor=blue!80!black, linkcolor=red!60!violet,urlcolor=blue!80!black]{hyperref}
\VerbatimFootnotes
\def\qed{\rule{2mm}{2mm}}

\usepackage{tikz,graphicx,pgfplots,pgfkeys,subcaption}   %The subcaption package is used in place of "subfig"
\usetikzlibrary{datavisualization.formats.functions}
\pgfplotsset{compat=1.12} %backwards compatibility 

\def\addlegendimage{\csname pgfplots@addlegendimage\endcsname}

\usepackage[pagewise,mathlines]{lineno}
\mathchardef\dash="2D
\newtheorem{theorem}{Theorem}[section]
\newtheorem{lemma}{Lemma}[section]

\newtheorem{assumption}{Assumption}[section]
\theoremstyle{definition}

\newtheorem{remark}{Remark}[section]
\usepackage{etoolbox} % This package goes here to add \qed to remarks automatically.
\AtEndEnvironment{remark}{~\qed}%

\DeclareMathOperator*{\argmax}{argmax}
\DeclareMathOperator*{\supp}{supp}

\begin{document}

\author{
Federico A. Bugni\\
Department of Economics\\
Duke University\\
\url{federico.bugni@duke.edu}
\and
Ivan A. Canay\\
Department of Economics\\
Northwestern University\\
\url{iacanay@northwestern.edu}
}

\title{Testing Continuity of a Density via \emph{g}-order statistics \\in the Regression Discontinuity Design\thanks{The research of the first author was supported by NSF Grant SES-1729280 and the research of the second author was supported by NSF Grant SES-1530534. Special thanks go out to Tim Armstrong for pointing out a problem in Section \ref{sec:large-q} of an earlier manuscript. We also thank Xiaohong Chen, the Associate Editor, and three referees for helpful comments. We finally thank Jackson Bunting, Joe Long, and Deborah Kim for excellent research assistance.}}

\maketitle
\vspace{-1.5cm}
\begin{abstract}
In the regression discontinuity design (RDD), it is common practice to assess the credibility of the design by testing the continuity of the density of the running variable at the cut-off, e.g., \cite{mccrary:08}. In this paper we propose an approximate sign test for continuity of a density at a point based on the so-called $g$-order statistics, and study its properties under two complementary asymptotic frameworks. In the first asymptotic framework, the number $q$ of observations local to the cut-off is \emph{fixed} as the sample size $n$ diverges to infinity, while in the second framework $q$ diverges to infinity slowly as $n$ diverges to infinity. Under both of these frameworks, we show that the test we propose is asymptotically valid in the sense that it has limiting rejection probability under the null hypothesis not exceeding the nominal level. More importantly, the test is easy to implement, asymptotically valid under weaker conditions than those used by competing methods, and exhibits finite sample validity under stronger conditions than those needed for its asymptotic validity. In a simulation study, we find that the approximate sign test provides good control of the rejection probability under the null hypothesis while remaining competitive under the alternative hypothesis. We finally apply our test to the design in \cite{lee:08}, a well-known application of the RDD to study incumbency advantage.
\end{abstract}

\thispagestyle{empty}

\noindent KEYWORDS: Regression discontinuity design, $g$-ordered statistics, sign tests, continuity, density. 

\noindent JEL classification codes: C12, C14.

\newpage
\setcounter{page}{1}

\section{Introduction}

The regression discontinuity design (RDD) has been extensively used in recent years to retrieve causal treatment effects - see \cite{lee/lemieux:10} and \cite{imbens/lemieux:08} for exhaustive surveys. The design is distinguished by its unique treatment assignment rule where individuals receive treatment when an observed covariate, known as the running variable, crosses a known cut-off. Such an assignment rule allows nonparametric identification of the average treatment effect (ATE) at the cut-off, provided that potential outcomes have continuous conditional expectations at the cut-off \citep[][]{hahn/etal:01}. The credibility of this identification strategy along with the abundance of such discontinuous rules have made RDD increasingly popular in empirical applications.

While the continuity assumption that is necessary for nonparametric identification of the ATE at the cut-off is fundamentally untestable, researchers routinely assess the plausibility of their RDD by exploiting two testable implications of a stronger identification assumption proposed by \cite{lee:08}. We can describe the two implications as follows: (i) the treatment is locally randomized at the cut-off, which translates into the distribution of all observed baseline covariates being continuous at the cut-off; and (ii) individuals have imprecise control over the running variable, which translates into the density of the running variable being continuous at the cut-off. The practice of judging the reliability of RDD applications by assessing either of the two above stated implications (commonly referred to as manipulation, or falsification, or placebo tests) is ubiquitous in the empirical literature. Indeed, Table \ref{tab:survey} surveys RDD empirical papers in four leading applied economic journals during the period 2011-2015. Out of 62 papers, 43 of them include some form of manipulation, falsification, or placebo test. \label{page:lee}

This paper proposes an approximate sign test for the null hypothesis on the second testable implication, i.e., the density of the running variable is continuous at the cut-off.\footnote{It is important to emphasize that the null hypothesis we test in this paper is neither necessary nor sufficient for identification of the ATE at the cut-off; see Remark \ref{rem:nec-suf}.} 
The approximate sign test has a number of distinctive attractive properties relative to existing methods used to test our null hypothesis of interest. First, the test does not require consistent non-parametric estimators of densities and simply exploits the fact that a certain functional of order statistics of the data is approximately binomially distributed under the null hypothesis. Second, our test controls the limiting null rejection probability under fairly mild conditions that, in particular, do not require existence of derivatives of the density of the running variable.\footnote{We use the term null rejection probability as opposed to asymptotic size as a way to acknowledge that, for a given sample size, there always exists a heavily steep smooth function that is indistinguishable from a discontinuous one. Remark \ref{rem:uniformity} discusses this further and provides important references.} In addition, our test is valid in finite samples under stronger, yet plausible, conditions. Third, the asymptotic validity of our test holds under two alternative asymptotic frameworks; one in which the number $q$ of observations local to the cut-off is \emph{fixed} as the sample size $n$ diverges to infinity, and one where $q$ diverges to infinity slowly as $n$ diverges to infinity. Importantly, both frameworks require similar and arguably mild conditions. Fourth, our test is simple to implement as it only involves computing order statistics, a constant critical value, and a single tuning parameter. This contrasts with existing alternatives that require local polynomial estimation of some order and either bias correction or under-smoothed bandwidth choices. Finally, we have developed a companion \verb+Stata+ package to facilitate the adoption of our test.\footnote{The \verb+Stata+ package \verb+rdcont+ can be downloaded from \url{http://sites.northwestern.edu/iac879/software/}.} 

The construction of our test is based on the simple intuition that, when the density of the running variable is continuous at the cut-off, the fraction of units under treatment and control local to the cut-off should be roughly the same. This means that the number of treated units out of the $q$ observations closest to the cut-off, is approximately distributed as a binomial random variable with sample size $q$ and probability $\frac{1}{2}$. To formalize this intuition, we exploit and develop properties of the so-called $g$-order statistics \citep[see, e.g.,][]{kaufmann/reiss:92,reiss:89} and consider the two asymptotic frameworks mentioned earlier to capture the local behavior of the density at the cut-off. In the first asymptotic framework, $q$ is fixed as $n\to \infty$ to represent a finite sample situation where the effective number of observations used by the test is too small to credibly invoke approximations for ``large'' $q$. This may arise, for example, when the density is not so well behaved around the cut-off as illustrated in some of our simulations. This framework is similar to the one in \cite{canay/kamat:18}, who in turn exploit results from \cite*{canay/romano/shaikh:17}. It is worth noting that the hypothesis we test, the test statistic, the critical value, and most of the formal arguments are different from those in \cite{canay/kamat:18} or \cite*{canay/romano/shaikh:17}. In the second asymptotic framework, $q$ diverges to infinity slowly as $n\to \infty$ to represent a finite sample situation where the effective number of observations used by the test is large enough to invoke approximations for ``large'' $q$. This framework is similar to the one in \cite{mccrary:08,otsu/etal:13,cattaneo/jansson/ma:17,armstrong/kolesar:19}, among others, and is in line with more traditional asymptotic arguments in non-parametric tests. 

From a technical standpoint, this paper has several contributions relative to the existing literature. To start, our results exhibit two important differences relative to \cite{canay/kamat:18} that go beyond the difference in the null hypotheses. First, we do not study our test as an approximate randomization test but rather as an approximate sign test. This not only requires different analytical tools, but also by-passes some of the challenges that would arise if we were to characterize our test as an approximate randomization test; see Remark \ref{rem:approx-rand} for a discussion on this. In addition, our approach in turn facilitates the analysis for the second asymptotic framework in which $q\to \infty$. Second, we develop results on $g$-order statistics as important intermediate steps towards our main results. Some of them may be of independent interest; e.g., Theorem \ref{thm:g-order-stats}. In addition, relative to the results in \citet[][]{mccrary:08,otsu/etal:13,cattaneo/jansson/ma:17}; our test does not involve consistent estimators of density functions to either side of the cut-off and does not require conditions involving existence of derivatives of the density of the running variable local to the cut-off. To the best of our knowledge, the formal asymptotic results we present are original to this paper. 

It is relevant to note that similar binomial tests have been recently proposed in the RDD literature by \cite{cattaneo/Tit/VB:16,cattaneo/Tit/VB:17} and \cite{frandsen:17}. As we explain in more detail in Remark \ref{rem:matias}, there are important differences between these binomial tests and ours when it comes to the null hypothesis being tested, the formal arguments, and the practical implementation of the tests. \cite{cattaneo/Tit/VB:16,cattaneo/Tit/VB:17} rely on finite sample arguments to justify their test construction for the hypothesis of local randomization. \cite{frandsen:17} also relies on finite sample arguments to test the hypothesis of manipulation of a discretely distributed running variable. In contrast, we test the hypothesis that the density of the running variable is continuous at the cut-off. Our focus on this particular null hypothesis prevents us from invoking finite sample arguments at the level of generality we consider and leads us to study the asymptotic properties of the approximate sign test. Our analysis also guides how to choose $q$ in  data-dependent way and this, in turn, leads to a distinctive implementation of the test that we propose. 

The remainder of the paper is organized as follows. Section \ref{sec:setup} introduces the notation and describes the null hypothesis of interest. Section \ref{sec:ourtest} defines $g$-order statistics, formally describes the test we propose, and discusses all aspects related to its implementation including a data-dependent way of choosing $q$. Section \ref{sec:results} presents the main formal results of the paper, dividing those results according to the two alternative asymptotic frameworks we employ. In Section \ref{sec:simulations}, we examine the relevance of our asymptotic analysis for finite samples via a simulation study. Finally, Section \ref{sec:application} implements our test to reevaluate the validity of the design in \cite{lee:08} and Section \ref{sec:conclusion} concludes. The proofs of all results can be found in the Appendix.

\section{Setup and notation}\label{sec:setup}

Let $Y\in \mathbf R$ denote the observed outcome of interest for an individual or unit in the population and $A\in \{0,1\}$ denote an indicator for whether the unit is treated or not. Further denote by $Y(1)$ the potential outcome of the unit if treated and by $Y(0)$ the potential outcome if not treated. As usual, the observed outcome and potential outcomes are related to treatment assignment by the relationship 
\begin{equation} \label{eq:obsy}
Y = Y(1)A + Y(0)(1 - A)~.
\end{equation}
The treatment assignment in the (sharp) RDD follows a discontinuous rule,
\begin{equation*}
A = I\{Z \geq \bar{z}\}~,
\end{equation*}
where $Z\in \mathcal Z\equiv \supp(Z)$ is an observed scalar random variable known as the running variable and $\bar{z}$ is the known threshold or cut-off value. For convenience we normalize $\bar{z}=0$, which is without loss of generality as we can always redefine $Z$ as $Z-\bar{z}$. This treatment assignment rule allows us to identify the average treatment effect (ATE) at the cut-off; i.e., 
\begin{equation*}
  E[Y(1) - Y(0)|Z=0]~.  
\end{equation*}
In particular, \cite{hahn/etal:01} establish that identification of the ATE at the cut-off relies on the discontinuous treatment assignment rule and the assumption that   
\begin{equation}\label{eq:hahn01}
    E[Y(1) | Z=z]\quad \text{and}\quad E[Y(0) | Z=z] \quad \text{are both continuous in $z$ at }z=0~.  
\end{equation}
Reliability of the RDD thus depends on whether the mean outcome for units marginally below the cut-off identifies the true counterfactual for those marginally above the cut-off.

The continuity assumption in \eqref{eq:hahn01} is arguably weak, but fundamentally untestable. In practice, researchers routinely employ two specification checks in RDD that, in turn, are testable implications of a stronger sufficient condition proposed by \citet[][Condition 2b]{lee:08}. The first check involves testing whether the distribution of pre-determined characteristics (conditional on the running variable) is continuous at the cut-off. See \cite{shen/zhang:16} and \cite{canay/kamat:18} for a recent treatment of this problem. The second check involves testing the continuity of the density of the running variable at the cut-off, an idea proposed by \cite{mccrary:08}. This second check is particularly attractive in settings where pre-determined characteristics are not available or where these characteristics are likely to be unrelated to the outcome of interest. Formally, we can state the hypothesis testing problem for the second check as  
\begin{equation}\label{eq:null}
      H_0: f_Z^{+}(0) = f_Z^{-}(0)\quad \text{vs.} \quad H_1: f_Z^{+}(0) \neq f_Z^{-}(0)~, 
\end{equation}
where $f_Z^{+}(0)$ and $f_Z^{-}(0)$ are the one-sided limits of the probability density function of $Z$, i.e.,
\begin{equation}\label{eq:fplus-fminus}
  f_Z^{+}(0) \equiv \lim_{z\downarrow 0} f_Z(z) \quad \text{and}\quad  f_Z^{-}(0) \equiv \lim_{z\uparrow 0} f_Z(z)~.
\end{equation}
In RDD empirical studies, the aforementioned specification checks are often implemented (with different levels of formality) and referred to as falsification, manipulation, or placebo tests (see Table \ref{tab:survey} for a survey).

In this paper we consider an approximate sign test for the null hypothesis of continuity in the density of the running variable $Z$ at the cut-off $\bar{z}=0$, i.e., \eqref{eq:null}. This test has three attractive features compared to existing approaches \citep[see, e.g.,][]{mccrary:08,otsu/etal:13,cattaneo/jansson/ma:17}. First, it does not require commonly imposed smoothness conditions on the density of $Z$, as it does not involve non-parametric estimation of such a density. Second, it exhibits finite sample validity under certain (stronger) easy to interpret conditions. Finally, it involves a single tuning parameter \citep[a feature shared by the approach proposed by][]{cattaneo/jansson/ma:17} as opposed to multiple ones in \cite{mccrary:08}. We discuss these features further in Section \ref{sec:results}.

\begin{remark}\label{rem:rothe}
  \cite{gerard/etat:2016} study the consequences of discontinuities in the density of $Z$ at the cut-off. In particular, the authors consider a situation in which manipulation occurs only for a subset of participants and use  the magnitude of the discontinuity of $f(z)$ at $z=0$ to identify the proportion of always-assigned units among all units close to the cut-off. Using this setup, \cite{gerard/etat:2016} show that treatment effects in RDD are not point identified but the model still implies informative bounds.
\end{remark}

\begin{remark}\label{rem:nec-suf}
It is important to emphasize that a running variable with a continuous density is neither necessary nor sufficient for the identification of the average treatment effect at the cut-off. For a discussion of this and some intuitive examples, see \cite{lee:08} and \cite{mccrary:08}. 
\end{remark}

\section{Approximate sign test via \emph{g}-ordered statistics}\label{sec:ourtest}
Let $P$ be the distribution of $Z$ and $Z^{(n)}=\{Z_i:1\le i\le n\}$ be a random sample of $n$ i.i.d.\ observations from $P$. Let $q$ be a small (relative to $n$) positive integer and $g:\mathcal Z \to \mathbf R$ be a measurable function such that $g(Z)$ has a continuous distribution function. For any $z,z'\in \mathcal Z$ define $\le_{g}$ as
\begin{equation*}
	z\le_{g}z' \quad \text{ if }\quad g(z)\le g(z')~.
\end{equation*}
The ordering defined by $\le_{g}$ is called a $g$-ordering on $\mathcal Z$. The $g$-order statistics $Z_{g,(i)}$ corresponding to $Z^{(n)}$ are defined as the values satisfying 
\begin{equation*}
	Z_{g,(1)}\le_{g} \cdots \le_{g}Z_{g,(n)}~,
\end{equation*}
see, e.g., \citet[Section 2.1]{reiss:89} and \cite{kaufmann/reiss:92}. 

To construct our test statistic, we use the sign of the $q$ values of $\{Z_i:1\le i\le n\}$ that are induced by the $q$ smallest values of $\{g(Z_i)=|Z_i|:1\le i\le n\}$. That is, for $Z_{g,(1)},\dots,Z_{g,(q)}$, let 
\begin{equation}\label{eq:Agj}
  A_{g,(j)} \equiv I\{Z_{g,(j)}\ge 0\} \text{ for } 1\le j\le q~,
\end{equation}
and 
\begin{equation}\label{eq:Sn}
  S_n \equiv \sum_{j\le q} A_{g,(j)}~.
\end{equation}
The test statistic of our test only depends on the data via $S_n$ and is defined as 
\begin{equation}\label{eq:T}
	T_q(S_n) \equiv \sqrt{q}\left|\frac{1}{q}S_n -\frac{1}{2} \right|~.
\end{equation}

In order to describe the critical value of our test it is convenient to recall that the cumulative distribution function (CDF) of a binomial random variable with $q$ trials and probability of success $\frac{1}{2}$ is given by 
\begin{equation}\label{eq:binom-cdf}
  \Psi_q(b) \equiv \frac{1}{2^{q}}\sum_{x=0}^{\lfloor b\rfloor }\binom{q}{x}I\{0 \leq b \leq q\} ~+~ I\{b > q\}~,
\end{equation}
where $\lfloor x \rfloor$ is the largest integer not exceeding $x$. Using this notation the critical value for a significance level $\alpha\in (0,1)$ is given by
\begin{equation}\label{eq:cq}
  c_q(\alpha) \equiv \sqrt{q}\left(\frac{1}{2}-\frac{b_q(\alpha)}{q}\right)~,
\end{equation}
where $b_q(\alpha)$ is the unique value in $\{0,1,\ldots ,\lfloor \frac{q}{2}\rfloor \} $ satisfying  
\begin{equation}\label{eq:bq}
  \Psi_q(b_q(\alpha)-1)\le \frac{\alpha}{2} < \Psi_q(b_q(\alpha))~.
\end{equation}
The test we propose is then given by 
\begin{equation}\label{eq:testmain}
  \phi(S_n) = \begin{cases}
    1   & \text{ if } T_q(S_n)>c_q(\alpha)\\
    a_q(\alpha) & \text{ if } T_q(S_n)=c_q(\alpha)\\
    0   & \text{ if } T_q(S_n)<c_q(\alpha)
  \end{cases}~,
\end{equation}
where 
\begin{equation}\label{eq:aq}
  a_q(\alpha) \equiv 2^{q-1}\binom{q}{b_q(\alpha)}^{-1} \left[\alpha-2\Psi_q(b_q(\alpha)-1)\right]~.
\end{equation}

Intuitively, the test $\phi(S_n)$ exploits the fact that, under the null hypothesis in \eqref{eq:null}, the distribution of the treatment assignment should be locally the same to either side of the cut-off. That is, local to the cut-off, the treatment assignment behaves as purely randomized under the null hypothesis, so the fraction of units under treatment and control should be similar.  

\begin{remark}\label{rem:non-randomized}
  The test in \eqref{eq:testmain} is possibly randomized. The non-randomized version of the test takes the form $I\{T_q(S_n) > c_q(\alpha)\}$ and, by definition, does not reject more often than $\phi(S_n)$ in \eqref{eq:testmain}.  For our data-dependent choice of $q$ that we describe in the next section, the randomized and non-randomized versions perform similarly in our simulations.
\end{remark}

\begin{remark}\label{rem:unique}
  The value of $b_q(\alpha)\in \{0,1,\ldots ,\lfloor \frac{q}{2}\rfloor\}$ solving \eqref{eq:bq} is well-defined and unique for all $q\ge 1$ and $\alpha\in(0,1)$. To see this, let 
  \begin{equation}\label{eq:q-ast-alpha}
    q^{\ast}(\alpha) \equiv 1-\frac{\log \alpha}{\log 2}~.
  \end{equation}
  When $q<q^{\ast}(\alpha) $, \eqref{eq:bq} uniquely holds for $b_q(\alpha)=0$. In this case, $\phi(S_n)$ in \eqref{eq:testmain} does not reject deterministically with positive probability. When $q\ge q^{\ast}(\alpha)$, the uniqueness of the solution is guaranteed by $\Psi_q(\cdot)$ being strictly increasing over $\{0,1,\ldots ,\lfloor \frac{q}{2}\rfloor\}$, $\Psi_{q}(0)=\frac{1}{2^q}$, and $\Psi_{q}(\frac{q}{2})\ge \frac{1}{2}$. In this case, $\phi(S_n)$ in \eqref{eq:testmain} deterministically rejects with positive probability. This shows that in order for the non-randomized version of the test to be non-trivial (see Remark \ref{rem:non-randomized}), $q$ needs to exceed $q^*(\alpha)$. To better appreciate these magnitudes, note that for $\alpha=5\%$ this requires $q\ge 6$ while for $\alpha=1\%$ this requires $q\ge 8$. Similarly, and given $b_q(\alpha)$, the value of $a_q(\alpha)$ in \eqref{eq:aq} is also uniquely defined and taking values in $[0,1)$ by the same properties of $\Psi_q(\cdot)$. 
\end{remark}

Given $q$, the implementation of our test proceeds in the following five steps. 

\begin{itemize}
  \item[\bf Step 1.] Find the $q$ observations closest to the cut-off, i.e., $Z_{g,(1)},\dots,Z_{g,(q)}$.
  \item[\bf Step 2.] Count the number of non-negative observations in $Z_{g,(1)},\dots,Z_{g,(q)}$, i.e., $S_n$ as in \eqref{eq:Sn}.
  \item[\bf Step 3.] Compute test statistic $T_q(S_n)$ as in \eqref{eq:T}, $c_q(\alpha)$ as in \eqref{eq:cq}, and $a_q(\alpha)$ as in \eqref{eq:aq}. 
 \item[\bf Step 4.]  Compute the p-value of the non-randomized version of the test as
\begin{equation}\label{eq:p-value}
  p_{\rm value} = 2\min\left\lbrace \Psi_q\left(S_n\right),\Psi_q\left(q-S_n\right)\right\rbrace ~. 
\end{equation}
  \item[\bf Step 5.] Reject the null hypothesis in \eqref{eq:null} using $\phi(S_n)$ in \eqref{eq:testmain}. If a non-randomized test is preferred, reject the null hypothesis if $p_{\rm value}<\alpha$. 
\end{itemize}

\begin{remark}\label{rem:matias}
As we show in Theorems \ref{thm:g-order-stats} and \ref{thm:main}, the test $\phi(S_n)$ is an approximate sign or binomial test. As mentioned in the introduction, related binomial tests have been recently presented in the RDD context by \cite{cattaneo/Tit/VB:16}, \cite{cattaneo/Tit/VB:17}, and \cite{frandsen:17}. The first two papers use a binomial test based on the number of observations of the running variable exceeding the cut-off in a window $[-h, h]$ for a varying bandwidth $h$. The authors propose to vary $h$ until a ``break-down'' window size $h^*$ is found, which is defined as the largest window such that the minimum $p$-value of the binomial test is larger than $\alpha$ for all nested (smaller) windows. The justification provided for the validity of such a test involves a finite sample argument: under the hypothesis of ``local randomization/random assignment'' in $[-h^*, h^*]$, a binomial test with probability $\pi$ is exact. \cite{frandsen:17} considers an RDD model in which the running variable is discretely distributed and tests a different hypothesis from ours. Also motivated by finite sample arguments, he proposes a test that involves quantiles from binomial distributions. Contrary to these papers, our goal is not to validate a ``local random assignment'' hypothesis or to deal with discrete running variables in an RDD framework, but rather to test the continuity hypothesis in \eqref{eq:null} when the running variable is continuous at the cut-off. As a result of this, we cannot exploit finite sample arguments and rather need to rely on the asymptotic analysis of our test. The formal results in Theorems \ref{thm:g-order-stats}, \ref{thm:main}, and \ref{thm:largeq-size} are novel to this paper and, to the best of our knowledge, they provide the first formal results about approximate sign tests for the hypothesis in \eqref{eq:null} in the RDD framework.
\end{remark}

\subsection{Data-dependent rule for \emph{q}}\label{sec:q-rot}
In this section we discuss the practical considerations involved in the implementation of our test, highlighting how we addressed these considerations in the companion \verb+Stata+ package. The only tuning parameter of our test is the number $q$ of observations closest to the cut-off. We propose a data-dependent way to choose $q$ that combines a rule of thumb with a local optimization. We call this data-dependent rule the ``informed rule of thumb'' and its computation requires the two steps described below. For the sake of clarity, in this section we do not use the normalization $\bar{z}=0$. Additional computational details are presented in Appendix \ref{app:q-irot}.

In Section \ref{sec:large-q} we consider the asymptotic framework where $q$ diverges as $n\to\infty$. Under Assumption \ref{ass:A} and $H_0$ in \eqref{eq:null}, we show in that section that the value of $q$ that sets the worst case asymptotic bias equal to the standard deviation is given by
\begin{equation}\label{eq:q-intuition}
     q = n^{2/3}\left(\frac{4f^2_Z(\bar{z})}{C_P}\right)^{2/3} ~, 
\end{equation}
where $f_Z(\bar{z})$ equals $f_Z^{+}(\bar{z})=f_Z^{-}(\bar{z})$ under $H_0$, and $C_P$ is the Lipschitz constant in Assumption \ref{ass:A}(i'). Since the results in Theorem \ref{thm:largeq-size} also require $q^{3/2}/n\to 0$, we propose to start with an initial rule of thumb where $f_Z(\bar{z})$ and $C_P$ are computed under the assumption $Z\sim N(\mu, \sigma^2)$ and the rate is set to $n^{1/2}$. This leads to    
\begin{equation}\label{eq:qrot}
    q_{\rm rot} = n^{1/2} \left( \sigma \frac{4\phi^2_{\mu,\sigma}(\bar{z})}{\phi_{\mu,\sigma}(\mu+\sigma)}\right)^{2/3}~, 
\end{equation}
where we used that $C_P = |\phi'_{\mu,\sigma}(\mu+\sigma)| = \frac{1}{\sigma}\phi_{\mu,\sigma}(\mu+\sigma)$ when $Z\sim N(\mu,\sigma^2)$, and  $\phi_{\mu,\sigma}(\cdot)$ and $\phi'_{\mu,\sigma}(\cdot)$ denote the density of $N(\mu,\sigma^2)$ and its derivative. This initial rule of thumb is location and scale invariant and, by definition, is inversely related to the asymptotic bias of the test statistic in the asymptotic framework of Section \ref{sec:large-q}. In turn, the constant multiplying $n^{1/2}$ in \eqref{eq:qrot} is fairly intuitive. First, it captures the idea that a steeper density at the cut-off should be associated with a smaller value of $q$. Intuitively, the steeper the density, the more it resembles a density that is discontinuous (Figure \ref{fig:Designs}.(c) illustrates this in Section \ref{sec:simulations}). Since the maximum slope is determined by the Lipschitz constant, the rule is inversely proportional to that. Second, it also captures the idea that $q$ should be small if the cut-off is a point of low density. Intuitively, when $f_Z(\bar z)$ is low, the $q$ closest observations to $\bar{z}$ are likely to be ``far'' from $\bar z$ (Figure \ref{fig:Designs}.(a) with $\mu=-2$ illustrates this in Section \ref{sec:simulations}). One could alternatively replace the normality assumption with a non-parametric estimator of $f_Z(\bar{z})$ but it is unfortunately impossible to choose $C_P$ adaptively for testing \eqref{eq:null} \citep[see, e.g.,][]{low:97,armstrong/kolesar:18}. Since any data-dependent rule for $q$ will require a reference for $C_P$, we prefer to prioritize its simplicity and use normality for both $f_Z(\bar{z})$ and $C_P$. 

The second step involves a local maximization of the asymptotic null rejection probability of the non-randomized version of the test. In particular, based on our results, we propose 
 \begin{equation}\label{eq:q-irot}
    q_{\rm irot}=\argmax_{q \in \mathcal N(q_{\rm rot})} \Psi_q(b_q(\alpha)-1)~, 
  \end{equation}
where $\Psi_q(\cdot)$ is the CDF defined in \eqref{eq:binom-cdf}, $b_q(\alpha)$ is defined in \eqref{eq:bq}, and $\mathcal N(q_{\rm rot})$ is a discrete neighborhood defined in \eqref{eq:ENE} in Appendix \ref{app:q-irot}. This step helps the performance of the non-randomized version of the test (see Remark \ref{rem:non-randomized}) as $ \Psi_q(b_q(\alpha)-1)$ is non-monotonic in $q$ (see Figure \ref{fig:q-informed}) and so optimizing locally to $q_{\rm rot}$ over $\mathcal N(q_{\rm rot})$ prevents choosing a value of $q$ with a low value of $\Psi_q(b_q(\alpha)-1)$. In practice, we replace $(\mu,\sigma)$ with sample analogs to obtain the feasible informed rule of thumb $\hat q_{\rm irot}$.

\begin{remark}\label{rem:optimal-q}
  The recommended choice of $q$ in \eqref{eq:q-irot} can be interpreted as the under-smoothed version of the rule that captures a bias-variance trade-off, where we impose normality to compute unknown constants. It also exploits the shape of the limiting null rejection probability of the non-randomized version of the test to derive a better choice of $q$. Even though $q_{\rm irot}$ is motivated by a root-mean-square error (RMSE) optimal choice of $q$, it is an under-smoothed rule of thumb that is not optimal in a formal sense. A formal study of an optimal choice of $q$ in either of our asymptotic frameworks is an important topic of investigation that we leave for future research. 
\end{remark}

\section{Asymptotic framework and formal results}\label{sec:results}
In this section we derive the asymptotic properties of the test in \eqref{eq:testmain} using two alternative asymptotic frameworks. The first one requires $q$ to be fixed as $n\to \infty$, and represents a finite sample situation where the effective number of observations used by the test is too small to credibly invoke approximations for ``large'' $q$. The second framework requires $q\to \infty$ slowly as $n\to \infty$, and represents a finite sample situation where the effective number of observations used by the test is large enough to invoke approximations for ``large'' $q$.

There are three main features of our results that are worth highlighting: (i) our test exhibits similar properties under both asymptotic frameworks, (ii) the implementation of the test \emph{does not} depend on which asymptotic framework one has in mind, and (iii) all formal results require similar, and arguably mild, conditions. We start by introducing these conditions. 

\begin{assumption}\label{ass:A} The distribution function $P$ is absolutely continuous on $(-\delta,\delta)$ for some $\delta>0$. On this set, the density function $f_Z(z)$ satisfies the following:
\vspace{-2mm}
  \begin{enumerate}
    \item[(i)] $f_Z(z)$ is bounded on $(-\delta,\delta)$ and has one-sided limits at zero given by $f^{+}_{Z}(0)$ and $f^{-}_{Z}(0)$.
    \item[(i')] $\exists C_P\in(0,\infty)$ such that
    $$|f_{Z}(z)-f^{+}_{Z}(0)|\le C_P|z| \text{ for } z\in(0,\delta) \text{ and }  |f_{Z}(z)-f^{-}_{Z}(0)|\le C_P|z| \text{ for } z\in(-\delta,0)~.$$
    \item[(ii)] $f^{-}_{Z}(0)+f^{+}_{Z}(0)>0$.
  \end{enumerate}
\end{assumption} 

Assumptions \ref{ass:A}(i) and \ref{ass:A}(i') each impose different degrees of smoothness on the density of $Z$ local to the cut-off $\bar{z}=0$. Indeed, Assumption \ref{ass:A}(i') strengthens Assumption \ref{ass:A}(i) by replacing the requirement of left- and right-continuity at the cut-off with its Lipschitz version. In the formal results that follow, we use Assumption \ref{ass:A}(i) in the asymptotic framework where $q$ is fixed as $n\to \infty$ and Assumption \ref{ass:A}(i') in the asymptotic framework where $q\to \infty$ as $n\to \infty$. Both assumptions allow for the distribution of $Z$ to be discontinuous outside of a neighborhood of the cut-off.\footnote{In Appendix \ref{app:discontinuity} we also allow for situations with a mass point at the cut-off, i.e., $P\{Z=0\}>0$.}  More importantly, they do not require the density of $Z$ to be differentiable anywhere. This is in contrast to \cite{mccrary:08}, who requires three continuous and bounded derivatives of the density of $Z$ (everywhere except possibly at $\bar{z}=0$), and \cite{cattaneo/jansson/ma:17} and \cite{otsu/etal:13}, who require the density of $Z$ to be twice continuously differentiable local to the cut-off (in the case of a local-quadratic approximation). Assumption \ref{ass:A}(ii) rules out a situation where $f^{-}_{Z}(0)=f^{+}_{Z}(0)=0$, which is implicitly assumed away in \cite{mccrary:08} and \cite{otsu/etal:13}, and is weaker than assuming a positive density of $Z$ in a neighborhood of the cut-off as in \cite{cattaneo/jansson/ma:17}. In Section \ref{sec:simulations} we explore the sensitivity of our results to violations of these conditions. 

\subsection{Results for fixed \emph{q}}\label{sec:fixed-q}
In this section we present two main results. The first result, Theorem \ref{thm:g-order-stats}, describes the asymptotic properties of $S_n$ in \eqref{eq:Sn} when $q$ is fixed as $n\to\infty$. This result about $g$-order statistics with $g(\cdot)=|\cdot|$ represents an important milestone in proving the asymptotic validity of our test. The second result, Theorem \ref{thm:main}, exploits Theorem \ref{thm:g-order-stats} to show that the test in \eqref{eq:testmain} controls the limiting rejection probability under the null hypothesis. 

\begin{theorem}\label{thm:g-order-stats}
  Let Assumptions \ref{ass:A}(i) and \ref{ass:A}(ii) hold and let $q\geq 1$ be fixed. Then,
  \begin{equation*}
    S_n \overset{d}{\to} S\sim {\rm Bi}(q,\pi_f)
  \end{equation*}
  as $n\to \infty$, where ${\rm Bi}(q,\pi_f)$ denotes the Binomial distribution with $ q$ trials and probability of success 
  \begin{equation*}
    \pi_f\equiv  \frac{f^{+}_{Z}(0)}{f^{-}_{Z}(0)+f^{+}_{Z}(0)} ~. 
  \end{equation*} 
\end{theorem}

Theorem \ref{thm:g-order-stats}, although fairly intuitive, does not follow from standard arguments. First, the random variables $\{A_{g,(j)}:1\le j\le q\}$ are indicators of $g$-order statistics so, in general, they are neither independent nor identically distributed. Second, applying results from the literature on $g$-order statistics \cite[e.g.,][Theorem 1]{kaufmann/reiss:92} requires $g(Z)=|Z|$ to have a continuous distribution function everywhere on its domain. Under Assumption \ref{ass:A}(i) this is only true in $[0,\delta)$, and mass points are allowed outside of $[0,\delta)$. In the proof of Theorem \ref{thm:g-order-stats} we use a smoothing transformation of $Z$ as an intermediate step and then accommodate the results in \citet[][Theorem 1]{kaufmann/reiss:92} to reach the desired conclusion. 

The following result, which heavily relies on Theorem \ref{thm:g-order-stats}, is the main result of this section and characterizes the asymptotic properties of the test $\phi(S_n)$ in \eqref{eq:testmain}. 

\begin{theorem}\label{thm:main}
  Let Assumptions \ref{ass:A}(i) and \ref{ass:A}(ii) hold and let $q\ge 1$ be fixed. Then, the following holds for $\alpha\in(0,1)$:
\begin{enumerate}[(a)]
\item Under $H_0$ in \eqref{eq:null},
  \begin{equation*}
    \lim_{n\rightarrow \infty }E[\phi(S_n)]=2\Psi_q(b_q(\alpha)-1)+\frac{a_q(\alpha)}{2^{q-1}}\binom{q}{b_q(\alpha)}=\alpha~. 
  \end{equation*}
\item Under $H_1$ in \eqref{eq:null}, $\lim_{n\rightarrow \infty }E[\phi(S_n)] \geq \alpha$. 
\end{enumerate}
\end{theorem}

Theorem \ref{thm:main} shows that $\phi(S_n)$ behaves asymptotically, as $n\to \infty$, as the two-sided sign test in an experiment where one observes $S\sim {\rm Bi}(q,\pi)$ and wishes to test the hypotheses $H_0:\pi=\frac{1}{2}$ versus $H_1:\pi\ne \frac{1}{2}$. For this reason, we refer to $\phi(S_n)$ as an approximate sign test. 

\begin{remark}\label{rem:approx-rand}
  The test $\phi(S_n)$ could be alternatively characterized as an ``approximate'' randomization test, see \cite{canay/romano/shaikh:17} for a general description of such tests. However, such a characterization would make the analysis of the formal properties of the test more complicated and, in particular, the results in \cite{canay/romano/shaikh:17} would not immediately apply due to two fundamental challenges. First, Assumption 3.1(iii) in \cite{canay/romano/shaikh:17} is immediately violated in our setting. Second, such an approach would require an asymptotic approximation to the joint distribution of $\{A_{g,(j)}:1\le j\le q\}$, which in turn would require a strengthening of Lemma \ref{lemma:KR}. Our proof approach avoids both of these technicalities by directly exploiting the binary nature of $\{A_{g,(j)}:1\le j\le q\}$ and by simply approximating the distribution of $S_n$, which is a scalar, as in Theorem \ref{thm:g-order-stats}. 
\end{remark}

\begin{remark}\label{rem:finite-sample}
  It is possible to show that $\phi(S_n)$ in \eqref{eq:testmain} is level $\alpha$ in finite samples whenever the distribution of $Z$ is continuous and symmetric about the cut-off. In this case, the fundamental result in Lemma \ref{lemma:KR} holds for $S_n$ with $P\{Z>0~|~|Z|< r\}=\frac{1}{2}$ for any $r>0$, and the proof of Theorem \ref{thm:main} can in turn be properly modified to show $E[\phi(S_n)]=\alpha$ for all $n\ge 1$. 
\end{remark}

\subsection{Results for large \emph{q}}\label{sec:large-q}
In this section we study the properties of $\phi(S_n)$ in \eqref{eq:testmain} in an asymptotic framework where $q$ diverges to infinity as $n\to \infty$. This asymptotic framework is in line with traditional non-parametric arguments and so our results depend on the assumed smoothness of the density of $Z$ and the rate at which $q$ is allowed to grow. Importantly, the results in this section follow from Assumption \ref{ass:A}(i')-(ii) and so, accounting for the differences between Assumptions \ref{ass:A}(i) and \ref{ass:A}(i'), the result below shows that the asymptotic properties of the approximate sign test under both asymptotic frameworks require similar, and arguably mild, conditions. 

\begin{theorem}\label{thm:largeq-size}
Let Assumptions \ref{ass:A}(i') and \ref{ass:A}(ii) hold and let $q$ be such that $q\to\infty$ and  $\frac{q^{3/2}}{n}\to 0$ as $n\to\infty$. Then,
\begin{equation}\label{eq:main-thm4.3}
  \sqrt{q}\left(\frac{1}{q}S_n -\pi_f \right)\overset{d}{\to}N\left(0,\pi_f(1-\pi_f)\right)~,
\end{equation}
where $\pi_f$ is as in Theorem \ref{thm:g-order-stats}. Moreover, the following holds for $\alpha\in(0,1)$:
\begin{enumerate}[(a)]
\item Under $H_0$ in \eqref{eq:null}, $\lim_{n\rightarrow \infty }E[\phi(S_n)]=\alpha$.
\item Under $H_1$ in \eqref{eq:null}, $\lim_{n\rightarrow \infty }E[\phi(S_n)]=1$.
\item Under a sequence of alternative distributions local to $H_0$ satisfying $\sqrt{q}(\pi_f-\frac{1}{2})\to \Delta\ne 0$,
\begin{equation*}
   \lim_{n\rightarrow \infty }E[\phi(S_n)]=P\{|\zeta+2\Delta|>z_{\alpha/2}\}>\alpha~,
\end{equation*}
where $\zeta\sim N(0,1)$ and $z_{\alpha/2}$ is the $(1-\frac{\alpha}{2})$-quantile of $\zeta$.
\end{enumerate}
\end{theorem}

Theorem \ref{thm:largeq-size}, although fairly intuitive again, does not follow from standard arguments. In particular, given that the random variables $\{A_{g,(j)}:1\le j\le q\}$ are neither independent nor identically distributed, the result does not follow from a simple application of the central limit theorem. We instead adapt \citet[][Theorem 1]{kaufmann/reiss:92} and prove the result using first principles and the normal approximation to the binomial distribution. 

Given the result in Theorem \ref{thm:largeq-size}, we can provide some insight on the properties of the data-dependent rule for choosing $q$ that we describe in Section \ref{sec:q-rot}. Specifically, we focus on providing interpretation to $q_{\rm rot}$ in \eqref{eq:qrot}, as $q_{\rm irot}$ in \eqref{eq:q-irot} is a modification of $q_{\rm rot}$ to improve the performance of the non-randomized version of the test. Under $H_0$ in \eqref{eq:null} and Assumption \ref{ass:A}(i')-(ii), the results in \cite{armstrong/kolesar:19} imply that 
\begin{equation}\label{eq:expansion}
  \frac{\sqrt{q}(\frac{1}{q}S_n -\pi_f)}{\sqrt{\pi_f(1-\pi_f})}= \zeta_n + B_{n,q} + o_p(1)~,    
\end{equation}
where $\zeta_n \overset{d}{\to} \zeta \sim N(0,1)$ and $B_{n,q}$ is a standardized bias term satisfying 
\begin{equation}\label{eq:worst-bias}
  |B_{n,q}|\le \frac{q^{3/2}}{n}\frac{C_P}{4f^2_Z(0)}~   
\end{equation}
with $f_Z(0)$ equals $f_{Z}^{+}(0)=f_{Z}^{-}(0)$ under $H_0$. Denote by $t^{\ast}$ the right-hand side of \eqref{eq:worst-bias} and note that this can be interpreted as the worst (in absolute value) ratio of bias to standard deviation (sd) of the left-hand side of \eqref{eq:expansion}. We can then solve for $q$ to obtain
\begin{equation}\label{q:ast}
 	q^{\ast} = n^{2/3}(t^{\ast})^{2/3} \left(\frac{4f^2_Z(0)}{C_P}\right)^{2/3} ~.
\end{equation}
This derivation shows that the requirement $\frac{q^{3/2}}{n}\to 0$ is analogous to under-smoothing as this is the rate condition that removes the worst-case asymptotic bias.\footnote{A previous version of this paper did not include Assumption \ref{ass:A}(i') and the requirement $q^{3/2}/n\to 0$, which is required to control the asymptotic bias term. We thank Tim Armstrong for pointing this out to us.} This immediately gives two alternative interpretations to the data-dependent rule $q_{\rm rot}$ in \eqref{eq:qrot} \cite[see][Section 4.2, for a more detailed description]{armstrong/kolesar:19}. In order to describe these two interpretations, note that by \eqref{q:ast} and \eqref{eq:qrot} we obtain that $q_{\rm rot}=q^*$ whenever 
\begin{equation}\label{eq:t-ast}
    t^{\ast} =  \left(\frac{n^{1/2}}{n^{2/3}} \right)^{3/2}\left[\frac{\phi^2_{\mu,\sigma}(0)}{f^2_Z(0)}\frac{C_P}{\frac{1}{\sigma}\phi_{\mu,\sigma}(\mu+\sigma)} \right]~.
\end{equation}
Assume for a moment that the rule-of-thumb assumption of normality is correct (which means that the term within brackets in \eqref{eq:t-ast} equals 1). Then, $q_{\rm rot}$ is equivalent to $q^{\ast}$ for a worst ratio of bias to sd $t^{\ast}$ given by 
\begin{equation*}
	t^{\ast} = \left(\frac{n^{1/2}}{n^{2/3}} \right)^{3/2} \quad \Rightarrow\quad  t^{\ast} = 0.12 \text{ for } n=5,000~.
\end{equation*}
This implies the size of $\phi(S_n)$ for $\alpha=5\%$ and $n=5,000$ would approximately be $P\{|\zeta+0.12|>z_{\alpha/2}\} =5.16\%$. In this sense, $q_{\rm rot}$ makes the size distortion of the bias negligible when $n=5,000$. Next, suppose that the rule-of-thumb assumption of normality over-estimates the ratio $f^2_Z(0)/C_P$. In other words, suppose that the term within brackets in \eqref{eq:t-ast} equals a constant $a>1$. In this case, $q_{\rm rot}$ would be equivalent to $q^{\ast}$ for a worst ratio of bias to sd $t^{\ast}$ given by
\begin{equation*}
	t^{\ast} = a\left(\frac{n^{1/2}}{n^{2/3}} \right)^{3/2} \quad \Rightarrow\quad  t^{\ast} = 0.36 \text{ for } n=5,000 \text{ and }  a=3~.
\end{equation*}
This implies that the size of $\phi(S_n)$ for $\alpha=5\%$ and $n=5,000$ would approximately be $P\{|\zeta+0.36|>z_{\alpha/2}\} =6.38\%$. When $f_Z(0)=\phi_{\mu,\sigma}(0)$, this means that even if the true Lipschitz constant $C_P$ is three times larger than the one imposed by normality, $\phi(S_n)$ would still exhibit mild over-rejection under the null hypothesis. The price we pay for this robustness under the null hypothesis (in terms of performance and mild requirements) is possibly a lower power under the alternative hypothesis, a feature that we explore in the simulations of Section \ref{sec:simulations}. 

\begin{remark}\label{rem:normal-cv}
It may be tempting to use the first part of Theorem \ref{thm:largeq-size} to consider a variation of the test we propose; namely the test that rejects $H_0$ when $T_q(S_n)>\frac{1}{2}z_{\alpha/2}$ and $z_{\alpha/2}$ is the $(1-\frac{\alpha}{2})$-quantile of a standard normal random variable. However, we do not recommend this variation as it provides no theoretical advantages over $\phi(S_n)$ in the asymptotic framework where $q\to\infty$, and it is not formally justified in the asymptotic framework where $q$ is fixed (in particular, such a variation will not inherit the finite sample properties discussed in Remark \ref{rem:finite-sample}). 
\end{remark}

\begin{remark}\label{eq:Wald-Stat}
As pointed out by a referee, in the asymptotic framework where $q\to\infty$, the test statistic $T_q(S_n)$ can be shown to be proportional to a Wald-type statistic $$W_n = |\hat f_Z(h_n)-\hat f_Z(-h_n) |~, $$ where $\hat f_Z(z)$ is a non-parametric kernel density estimator of $f_Z(z)$ implemented with a uniform on $[-1,1]$ kernel and bandwidth $h_n$. Under some conditions it will follow that $W_n$ is asymptotically normal and a test for $H_0$ could be constructed by using the quantile of a normal distribution (possibly by additionally estimating the asymptotic variance). One could go a step further and use the bound on the bias term $B_{n,q}$ to construct a test that explicitly accounts for the asymptotic bias of the test following the approach proposed by \cite{armstrong/kolesar:19}. However, the interpretation of $\phi(S_n)$ as a test based on a Wald-type statistic with a normal critical value exclusively holds in the asymptotic framework where $q\to\infty$ and does not apply in the asymptotic framework with fixed $q$. For this reason, we do not emphasize this interpretation here.  
\end{remark}

\begin{remark}\label{rem:uniformity}
The recent literature has obtained impossibility results in the RDD setting that apply to the hypothesis testing problem in \eqref{eq:null}; see, e.g., \cite{low:97},  \cite{kamat:17}, \cite{armstrong/kolesar:18}, and \cite{bertanha/moreira:2019}. An implication of these impossibility results is that $\phi(S_n)$ cannot control size in a uniform sense without further restricting the set of data generating processes. These findings are reflected in the bound on the bias term presented in \eqref{eq:worst-bias}, where higher values of $C_P$ or lower values of $f_Z(0)$ can make such a bound arbitrarily high for given values of $q$ and $n$. We would therefore expect the performance of $\phi(S_n)$ to deteriorate in cases where the density at the cut-off is very low or very steep, as highlighted by the simulations we present next.  
\end{remark}

\section{Simulations}\label{sec:simulations}
In this section we examine the finite-sample performance of the test in \eqref{eq:testmain} with a simulation study. Instead of just presenting designs where this test excels relative to competing ones, we present an array of data generating processes that hopefully illustrate its relative strengths and weaknesses. The data for the study are simulated as i.i.d.\ samples from the following designs.

\begin{itemize}
\item[] \textbf{Design 1}: For $\mu\in\{-2,-1,0\}$, $Z \sim N(\mu,1)$.

\item[] \textbf{Design 2}: For $\lambda\in\{\frac{1}{3},1\}$, 
\begin{equation*}
  Z\sim \begin{cases}
  V_{1} & \text{with prob. } \lambda \\
  V_{2} & \text{with prob. } (1-\lambda)\\
  \end{cases} ~,
\end{equation*}
where $V_{1}\sim 2\text{Beta}(2,4) - 1$ and $V_{2}\sim 1-2\text{Beta}(2,8)$. 

\item[] \textbf{Design 3}: For $(\lambda_1,\lambda_2,\lambda_3)=(0.4,0.1,0.5)$, 
\begin{equation*}
  Z\sim \begin{cases}
  V_{1} & \text{with prob. } \lambda_1 \\
  V_{2} & \text{with prob. } \lambda_2\\
  V_{3} & \text{with prob. } \lambda_3\\
  \end{cases} ~,
\end{equation*}
where $V_1\sim N(-1,1)$, $V_2\sim N(-0.2,0.2)$, and $V_3\sim N(3,2.5)$.

\item[] \textbf{Design 4}: For $\kappa\in\{0.05,0.10,0.25\}$, the density of $Z$ is given by 
\begin{equation*}
  f_Z(z) = \begin{cases}
  0.75 & \text{if } z\in[-1,-\kappa]\\
  0.75 - \frac{1}{4\kappa}(z+\kappa)  & \text{if } z\in[-\kappa,\kappa]\\
  0.25 & \text{if } z\in[\kappa,1]
  \end{cases} ~. 
\end{equation*}

\item[] \textbf{Design 5}: For $\kappa\in\{0.05,0.10,0.25\}$, the density of $Z$ is given by 
\begin{equation*}
  f_Z(z) = \begin{cases}
  0.25 & \text{if } z\in[-1,-\kappa]\\
  0.50  & \text{if } z\in[-\kappa,\kappa]\\
  0.75 & \text{if } z\in[\kappa,1]
  \end{cases} ~. 
\end{equation*}

\item[] \textbf{Design 6}: We first non-parametrically estimate the density of the running variable in \citet[][see Section \ref{sec:application} for details]{lee:08} and then take i.i.d.\ draws from such a density. 
\end{itemize}

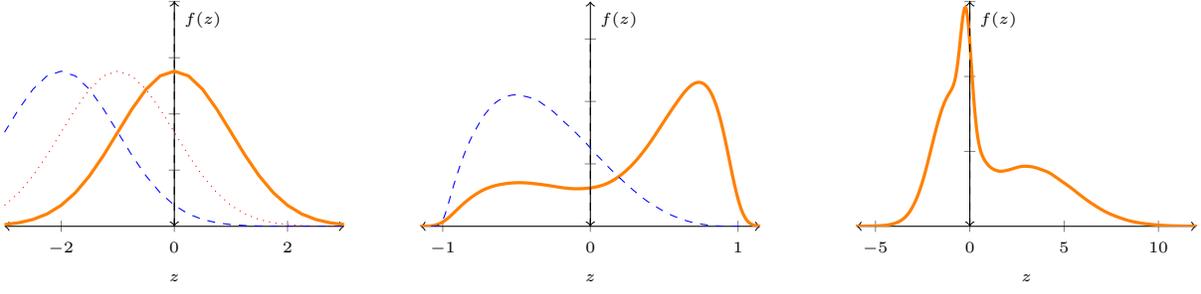
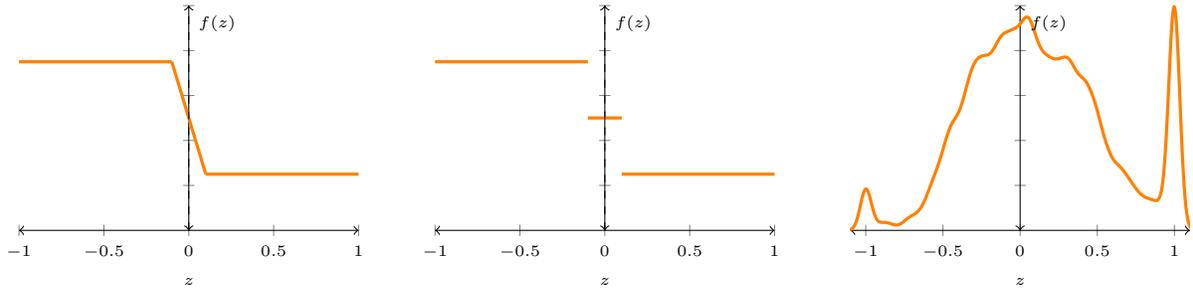
\begin{figure}[t!]
\centering
  %!TEX root = ../RDDcontinuity.tex
\pgfplotsset{every axis/.append style={
  axis x line=bottom,    % put the x axis in the middle
  axis y line=middle,    % put the y axis in the middle
  axis line style={<->}, % arrows on the axis
  label style={font=\tiny},
  tick label style={font=\tiny}  
},
cycle list={
{draw=black, solid,very thick,orange},
{draw=black, dashed,blue}, 
{draw=black,densely dashed,very thick,blue},
{draw=black, dashdotted, very thick,green},
{only marks, mark=asterisk}}}

\begin{subfigure}[t]{0.30\textwidth}\centering  
    \begin{tikzpicture}[baseline]
         \begin{axis}[width=2.4in,height = 1.8in,xmin=-3, xmax=3,ymin=0,ymax=0.8,xlabel=$z$,ylabel=$f(z)$,yticklabels={,,}]
              \addplot[domain=-3:3,very thick,orange] {1/(sqrt(3.28))*exp(-0.5*x^2)};
              \addplot[domain=-3:3,dotted,red] {1/(sqrt(3.28))*exp(-0.5*(x+1)^2)};
              \addplot[domain=-3:3,dashed,blue] {1/(sqrt(3.28))*exp(-0.5*(x+2)^2)};
              \draw[dashed] (0,0) -- (0,1.6);
         \end{axis}
    \end{tikzpicture}
    \caption{\footnotesize{Design 1: $\mu=(0,-1,-2)$ in solid, dotted, and dashed.}}\label{fig:D1}
\end{subfigure}\hfill
\begin{subfigure}[t]{0.33\textwidth}\centering  
    \begin{tikzpicture}[baseline]
        \begin{axis}[width=2.4in,height = 1.8in,xmin=-1.15, xmax=1.15,ymin=0,ymax=1.8,xlabel=$z$,ylabel=$f(z)$,yticklabels={,,}]
              \addplot  table[y = {"data.y"}]{Design2a.dat};
              \addplot  table[y = {"d.y"}]{Design2b.dat};
              \draw[dashed] (0,0) -- (0,1.5);
         \end{axis}
    \end{tikzpicture}
   \caption{\footnotesize{Design 2: $\lambda=(\frac{1}{3},1)$ in solid and dashed}}\label{fig:D2}
\end{subfigure}\hfill
\begin{subfigure}[t]{0.33\textwidth}\centering  
    \begin{tikzpicture}[baseline]
      \begin{axis}[width=2.4in,height = 1.8in,xmin=-6, xmax=12,ymin=0,ymax=0.3,xlabel=$z$,ylabel=$f(z)$,yticklabels={,,}]
              \addplot  table[y = {"data.y"}]{Design6.dat};
               \draw[dashed] (0,0) -- (0,1.5);
         \end{axis}
    \end{tikzpicture}
    \caption{\footnotesize{Design 3}}\label{fig:D3}
\end{subfigure}

\bigskip

% \begin{subfigure}[t]{0.3\textwidth}\centering  
%     \begin{tikzpicture}[baseline]
%          \begin{axis}[width=2.2in,height = 1.3in,xmin=-1, xmax=1,ymin=0,ymax=1.6,xlabel=$z$,ylabel=$f(z)$,yticklabels={,,}]
%               \addplot[domain=-1:0,very thick,orange] {2*(x+1)/2};
%               \addplot[domain=0:1,very thick,orange] {2*(1-x)/2};
%               \addplot[domain=-1/3:0,dashed,blue] {2*(x+1/3)/(4/9)};
%               \addplot[domain=0:1,dashed,blue] {2*(1-x)/(4/3)};
%               \draw[dashed] (0,0) -- (0,1.6);
%          \end{axis}
%     \end{tikzpicture}
%     \caption{design 3}\label{fig:D3}
% \end{subfigure}
\begin{subfigure}[t]{0.33\textwidth}\centering  
    \begin{tikzpicture}[baseline]
          \begin{axis}[width=2.4in,height = 1.8in,xmin=-1, xmax=1,ymin=0,ymax=1,xlabel=$z$,ylabel=$f(z)$,yticklabels={,,}]
              \addplot[domain=-1:-0.1,very thick,orange] {3/4};
              \addplot[domain=-0.1:0.1,very thick,orange] {3/4-(10/4)*(x+0.1)};
              \addplot[domain=0.1:1,very thick,orange] {1/4};
              \draw[dashed] (0,0) -- (0,1);
         \end{axis}
    \end{tikzpicture}
    \caption{\footnotesize{Design 4: $\kappa=0.1$}}\label{fig:D4}
\end{subfigure}\hfill
\begin{subfigure}[t]{0.33\textwidth}\centering  
    \begin{tikzpicture}[baseline]
       \begin{axis}[width=2.4in,height = 1.8in,xmin=-1, xmax=1,ymin=0,ymax=1,xlabel=$z$,ylabel=$f(z)$,yticklabels={,,}]
              \addplot[domain=-1:-0.1,very thick,orange] {3/4};
              \addplot[domain=-0.1:0.1,very thick,orange] {0.5};
              \addplot[domain=0.1:1,very thick,orange] {1/4};
              \draw[dashed] (0,0) -- (0,1);
         \end{axis}
    \end{tikzpicture}
    \caption{\footnotesize{Design 5: $\kappa=0.1$}}\label{fig:D5}
\end{subfigure}\hfill
\begin{subfigure}[t]{0.33\textwidth}\centering  
    \begin{tikzpicture}[baseline]
      \begin{axis}[width=2.4in,height = 1.8in,xmin=-1.1, xmax=1.1,ymin=0,ymax=1,xlabel=$z$,ylabel=$f(z)$,yticklabels={,,}]
              \addplot  table[y = {d.y}]{DesignLee.dat};
               %\draw[dashed] (0,0) -- (0,1.5);
         \end{axis}
    \end{tikzpicture}
    \caption{\footnotesize{Design 6}}\label{fig:D6}
\end{subfigure}
    \caption{\footnotesize{Density functions $f(z)$ for Designs 1 to 6 used in the Monte Carlo simulations}}
\label{fig:Designs}
\end{figure}

Design 1 in Figure \ref{fig:Designs}(a) is the canonical normal case and, by Remark \ref{rem:finite-sample}, our test is expected to control size in finite samples when $\mu=0$ but not when $\mu\in \{-2,-1\}$. Indeed, $\mu=-2$ is a challenging case due to the low probability of getting observations to the right of the cut-off. Design 2 in Figure \ref{fig:Designs}(b) is taken from \cite{canay/kamat:18}. Design 3 in Figure \ref{fig:Designs}(c) is a parametrization of the taxable income density in \citet[][Figure 8]{saez:10}. This design exhibits a spike (almost a kink) to the left of the cut-off which is essentially a violation of the smoothness assumptions required by \cite{mccrary:08} and \cite{cattaneo/jansson/ma:17}. It also exhibits a steep density at the cut-off, which also makes it a difficult case in general. Similar to Design 3, Design 4 in Figure \ref{fig:Designs}(d) also illustrates the difficulty in distinguishing a discontinuity from a very steep slope; see \cite{low:97}, \cite{kamat:17}, \cite{armstrong/kolesar:18}, and \cite{bertanha/moreira:2019} for a formal discussion. Here we can study the sensitivity to the slope by changing the value of $\kappa$. Design 5 in Figure \ref{fig:Designs}(e) requires $\delta$ in Assumption \ref{ass:A}(a) to be such that $\delta<\kappa$ in order for our approximations to be accurate, but as opposed to Design 4, it is locally symmetric around the cut-off. As $\kappa$ gets smaller, we expect our test to perform worse if $q$ is not chosen carefully. Finally, Design 6 in Figure \ref{fig:Designs}(f) draws data i.i.d.\ from the non-parametric density estimate of the running variable in \cite{lee:08}, i.e., $Z$ is the difference in vote shares between Democrats and Republicans.   

We consider sample sizes $n \in \{1,000; 5,000 \}$, a nominal level of $\alpha=10\%$, and perform $10,000$ Monte Carlo repetitions. Designs 1 to 6 satisfy the null hypothesis in \eqref{eq:null}. We additionally consider the same models under the alternative hypothesis by randomly changing the sign of observations in the interval $z\in [0,0.1]$ with probability $\Pr=0.2-2z$. We report results for the following tests.

\begin{itemize}

\item[] \textbf{AS-NR} and \textbf{AS-R}: the approximate sign test we propose in this paper in its two versions. The randomized version (AS-R) in \eqref{eq:testmain} and the non-randomized version (AS-NR) that rejects when $p_{\rm value}$ in \eqref{eq:p-value} is below $\alpha$, see Remark \ref{rem:non-randomized}. The tuning parameter $q$ is set to
\begin{equation*}\label{eq:fixed}
q \in \{ 20 , 50, 75, \hat{q}_{\rm irot} \}~,
\end{equation*}
where $\hat{q}_{\rm irot}$ is the feasible informed rule of thumb described in Section \ref{sec:ourtest} and Appendix \ref{app:q-irot}.

\item[] \textbf{McC}: the test proposed by \cite{mccrary:08}. We implement this test using the function \verb+DCdensity+ from the \verb+R+ package \verb+rdd+ (\verb+Ver+ 0.57), with the default choices for the bandwidth parameter and kernel type. 

\item[] \textbf{CJM}: the test proposed by \cite{cattaneo/jansson/ma:17}. We implement this test using the \Verb+rddensity+ function from the \verb+R+ package \verb+rddensity+ (\verb+Ver+ 1.0). We use jackknifed standard errors and bias correction, as these are the default choices in the paper.

\end{itemize}

\begin{table}[t!]
  \begin{center}
  \scalebox{0.85}{%!TEX root = ../RDDcontinuity.tex
{\small 
\begin{tabular}{l|ccccccccc|ccccccccc}
  \hline\hline
  \multicolumn{1}{c}{}  & \multicolumn{8}{c}{Rejection Rate under $H_0$}  & \multicolumn{1}{c}{} & & \multicolumn{8}{c}{Rejection Rate under $H_1$}\\ 
  \multicolumn{1}{c}{}  & \multicolumn{4}{c}{AS-NR} & & AS-R & McC & \multicolumn{1}{c}{CJM} & \multicolumn{1}{c}{} & & \multicolumn{4}{c}{AS-NR} & & AS-R & McC & CJM \\
  \multicolumn{1}{c}{}   & \multicolumn{4}{c}{$q$} & & $q$ &  &\multicolumn{1}{c}{} & \multicolumn{1}{c}{} & \multicolumn{1}{c}{} & \multicolumn{4}{c}{$q$} & & $q$ &  & \\
  \cline{2-5} \cline{7-7} \cline{12-15} \cline{17-17}
  \multicolumn{1}{c}{Design} & 20 & 50 & 75 & $\hat q_{\rm irot}$ & & $\hat q_{\rm irot}$ & &\multicolumn{1}{c}{} &\multicolumn{1}{c}{} &\multicolumn{1}{c}{} & 20 & 50 & 75 & $\hat q_{\rm irot}$ & & $\hat q_{\rm irot}$ & & \\  \hline
D1: $\mu=0$ &   4.4 &   6.8 &   6.6 & 10.0 &  & 10.1 &   9.2 &   8.2 &  &  &  11.1 &  19.6 &  18.1 & 25.2 &  & 25.4 &  17.0 &  11.4 \\ 
  D1: $\mu=-1$ &   4.3 &   8.1 &  12.4 & 10.5 &  & 10.6 &  11.9 &   9.4 &  &  &  10.7 &  21.5 &  26.3 & 24.8 &  & 24.9 &  21.2 &  10.4 \\ 
  D1: $\mu=-2$ &  12.4 &  84.5 &  99.8 &  8.3 &  & 11.3 &  11.4 &   7.6 &  &  &  17.2 &  87.8 &  99.9 & 12.0 &  & 15.4 &  11.8 &   7.8 \\ 
   \hline
D2: $\lambda=1$ &   4.0 &   7.0 &   7.9 & 10.4 &  & 10.6 &  11.2 &   9.6 &  &  &   9.4 &  13.4 &   9.7 & 19.5 &  & 19.7 &  26.4 &  16.8 \\ 
  D2: $\lambda=\frac{1}{3}$ &   4.2 &   7.0 &  10.3 & 10.6 &  & 10.7 &  10.7 &   7.6 &  &  &  11.9 &  32.0 &  42.6 & 32.1 &  & 32.3 &  34.6 &  18.2 \\ 
   \hline
D3 &   5.1 &  17.5 &  39.2 & 24.6 &  & 24.9 &  99.9 &  86.1 &  &  &  12.9 &  38.6 &  60.9 & 48.0 &  & 48.3 & 100.0 &  89.2 \\ 
   \hline
D4: $\kappa=0.25$ &   4.0 &   8.1 &  12.2 & 10.9 &  & 11.0 &  11.9 &   9.3 &  &  &  12.5 &  32.4 &  42.6 & 34.8 &  & 35.0 &  33.2 &  18.6 \\ 
  D4: $\kappa=0.10$ &   4.6 &  16.9 &  42.0 & 16.3 &  & 16.5 &  48.1 &  24.4 &  &  &  14.8 &  50.5 &  77.3 & 46.4 &  & 46.6 &  80.2 &  42.1 \\ 
  D4: $\kappa=0.05$ &   6.9 &  48.0 &  86.2 & 35.9 &  & 36.1 &  84.9 &  59.9 &  &  &  20.2 &  79.4 &  97.4 & 66.8 &  & 67.0 &  96.4 &  77.4 \\ 
   \hline
D5: $\kappa=0.25$ &   4.3 &   7.4 &   7.1 & 10.4 &  & 10.5 &  21.0 &  13.5 &  &  &  11.2 &  21.6 &  22.3 & 26.8 &  & 27.0 &  17.2 &  31.0 \\ 
  D5: $\kappa=0.10$ &   4.1 &   6.4 &   6.4 &  9.9 &  & 10.1 &  17.9 &  33.1 &  &  &  11.0 &  21.5 &  21.8 & 26.1 &  & 26.3 &  23.2 &  13.5 \\ 
  D5: $\kappa=0.05$ &   4.0 &   7.4 &  35.1 &  9.7 &  &  9.8 &  40.3 &  25.7 &  &  &  10.7 &  27.4 &  72.6 & 27.4 &  & 27.6 &  75.7 &  42.0 \\ 
   \hline
D6 &   4.2 &   6.2 &   6.6 &  9.4 &  &  9.6 &   9.5 &  10.6 &  &  &  11.5 &  25.1 &  27.8 & 32.8 &  & 33.0 &  37.0 &  23.8 \\ 
   \hline
\hline
\end{tabular}}
}
  \end{center}
  \caption{\footnotesize{Rejection probabilities (in $\%$) under $H_0$ and $H_1$ across Designs 1-6 and for $n=1,000$.}}
  \label{tab:T1000}
\end{table}

\begin{table}[t!]
  \begin{center}
  \scalebox{0.85}{%!TEX root = ../RDDcontinuity.tex
{\small 
\begin{tabular}{l|ccccccccc|ccccccccc}
  \hline\hline
  \multicolumn{1}{c}{}  & \multicolumn{8}{c}{Rejection Rate under $H_0$}  & \multicolumn{1}{c}{} & & \multicolumn{8}{c}{Rejection Rate under $H_1$}\\ 
  \multicolumn{1}{c}{}  & \multicolumn{4}{c}{AS-NR} & & AS-R & McC & \multicolumn{1}{c}{CJM} & \multicolumn{1}{c}{} & & \multicolumn{4}{c}{AS-NR} & & AS-R & McC & CJM \\
  \multicolumn{1}{c}{}   & \multicolumn{4}{c}{$q$} & & $q$ &  &\multicolumn{1}{c}{} & \multicolumn{1}{c}{} & \multicolumn{1}{c}{} & \multicolumn{4}{c}{$q$} & & $q$ &  & \\
  \cline{2-5} \cline{7-7} \cline{12-15} \cline{17-17}
  \multicolumn{1}{c}{Design} & 20 & 50 & 75 & $\hat q_{\rm irot}$ & & $\hat q_{\rm irot}$ & &\multicolumn{1}{c}{} &\multicolumn{1}{c}{} &\multicolumn{1}{c}{} & 20 & 50 & 75 & $\hat q_{\rm irot}$ & & $\hat q_{\rm irot}$ & & \\  \hline
D1: $\mu=0$ &   4.1 &   6.4 &   6.8 &  9.8 &  & 10.0 &   9.2 &   8.8 &  &  &  12.5 &  30.5 &  38.4 & 63.7 &  & 63.9 &  50.0 &  33.2 \\ 
  D1: $\mu=-1$ &   4.0 &   6.3 &   6.5 &  9.5 &  &  9.7 &  11.8 &   9.0 &  &  &  12.0 &  30.2 &  38.4 & 39.1 &  & 39.4 &  47.2 &  16.6 \\ 
  D1: $\mu=-2$ &   4.5 &  12.4 &  26.0 & 10.2 &  & 10.6 &  12.2 &   9.6 &  &  &  11.8 &  31.5 &  47.9 & 21.2 &  & 21.7 &  15.7 &  12.5 \\ 
   \hline
D2: $\lambda=1$ &   4.0 &   6.5 &   6.0 &  9.7 &  &  9.8 &   9.7 &   9.8 &  &  &  12.1 &  29.0 &  35.8 & 50.9 &  & 51.2 &  68.6 &  34.4 \\ 
  D2: $\lambda=\frac{1}{3}$ &   4.4 &   6.8 &   6.7 & 10.0 &  & 10.2 &  10.3 &   8.1 &  &  &  12.6 &  32.5 &  44.0 & 46.2 &  & 46.5 &  87.0 &  59.2 \\ 
   \hline
D3 &   4.2 &   7.2 &   7.9 & 17.2 &  & 17.4 & 100.0 &  93.9 &  &  &  12.4 &  33.6 &  46.6 & 73.7 &  & 73.9 & 100.0 &  97.8 \\ 
   \hline
D4: $\kappa=0.25$ &   4.7 &   6.3 &   6.4 & 11.2 &  & 11.4 &  11.6 &  10.2 &  &  &  13.5 &  33.8 &  45.5 & 69.9 &  & 70.2 &  78.8 &  61.5 \\ 
  D4: $\kappa=0.10$ &   4.2 &   7.1 &   8.1 & 16.9 &  & 17.0 &  68.6 &  22.8 &  &  &  13.3 &  36.3 &  51.5 & 80.0 &  & 80.2 &  99.5 &  72.4 \\ 
  D4: $\kappa=0.05$ &   4.2 &   8.3 &  12.0 & 36.7 &  & 36.9 &  99.7 &  93.5 &  &  &  14.5 &  42.0 &  62.1 & 91.9 &  & 92.0 & 100.0 &  99.7 \\ 
   \hline
D5: $\kappa=0.25$ &   4.0 &   6.2 &   6.0 &  9.7 &  &  9.8 &  33.4 &  17.0 &  &  &  12.1 &  30.8 &  40.8 & 60.1 &  & 60.4 &  55.2 &  78.5 \\ 
  D5: $\kappa=0.10$ &   4.0 &   6.3 &   6.4 & 10.0 &  & 10.2 &  50.9 &  43.9 &  &  &  12.6 &  31.3 &  40.9 & 60.8 &  & 61.1 &  33.2 &  13.3 \\ 
  D5: $\kappa=0.05$ &   4.3 &   6.6 &   6.9 & 10.5 &  & 10.7 &  45.7 &  30.9 &  &  &  12.8 &  31.7 &  40.3 & 60.8 &  & 61.1 &  98.2 &  63.7 \\ 
   \hline
D6 &   4.1 &   6.6 &   6.5 &  9.4 &  &  9.5 &  12.9 &  13.4 &  &  &  12.0 &  31.7 &  41.9 & 70.3 &  & 70.5 &  90.2 &  65.2 \\ 
   \hline
\hline
\end{tabular}}
}
  \end{center}
  \caption{\footnotesize{Rejection probabilities (in $\%$) under $H_0$ and $H_1$ across Designs 1-6 and for $n=5,000$.}}
  \label{tab:T5000}
\end{table}

\begin{table}[ht]
\hspace{-0.6cm}
\scalebox{0.80}{{\small 
\begin{tabular}{lccccccccccccccccc}
  \hline \hline
 & D1 & D1 & D1 & D2 & D2 & D3 & D4& D4& D4 & D5& D5& D5 & D6  \\ 
 &$\mu=0$ & $\mu=-1$ & $\mu=-2$ & $\lambda=\frac{1}{3}$ & $\lambda=1$ &  & $\kappa=0.25$& $\kappa=0.01$ & $\kappa=0.05$&  $\kappa=0.25$& $\kappa=0.01$ & $\kappa=0.05$& \\
  \hline
  &\multicolumn{13}{c}{$n=1,000$}\\
  \cline{2-14}
  AS-NR &   53.0 &   37.0 &    8.5 &   37.0 &   37.0 &   51.7 &   40.5 &   39.3 &   39.2 &   44.2 &   39.7 &   39.2 &   53.0 \\ 
  McC &  716.2 &  440.6 &  113.7 &  209.7 &  415.3 &  606.1 &  361.3 &  358.1 &  377.9 &  365.6 &  337.4 &  350.8 &  564.4 \\ 
  CJM &  574.7 &  434.1 &   85.1 &  142.8 &  456.6 &  493.7 &  351.4 &  412.8 &  437.9 &  359.2 &  338.5 &  416.9 &  498.8 \\ 
   \hline &\multicolumn{13}{c}{$n=5,000$}\\   \cline{2-14} AS-NR &  147.0 &   54.1 &   18.0 &  119.0 &   62.0 &  119.0 &  119.0 &  119.0 &  119.0 &  119.0 &  119.0 &  119.0 &  146.9 \\ 
  McC & 2964.3 & 1775.6 &  506.4 &  774.8 & 1782.3 & 2944.8 & 1526.3 & 1424.2 & 1612.0 & 1623.9 & 1318.4 & 1365.4 & 2258.3 \\ 
  CJM & 2312.5 & 1883.5 &  348.0 &  489.2 & 2007.7 & 1879.5 & 1263.9 & 1660.5 & 2076.4 & 1363.6 & 1194.7 & 1703.4 & 2274.9 \\ 
   \hline \hline
\end{tabular}
}}

 \caption{\footnotesize{Average effective sample size across simulations for each design. Effective sample size is defined as follows: $\hat q_{\rm irot}$ for AS-NR, number of obs.\ in $[-h_n,h_n]$ for McC, with $h_n$ being the bandwidth used to estimate the density to the left and to the right of the cutoff, and number of obs.\ in $[-h_{n,L},h_{n,R}]$ for CJM, with $h_{n,L}$ and $h_{n,R}$ being the bandwidths used to estimate the density the left and to the right of the cutoff, respectively.}}
  \label{tab:q-irot}
\end{table}

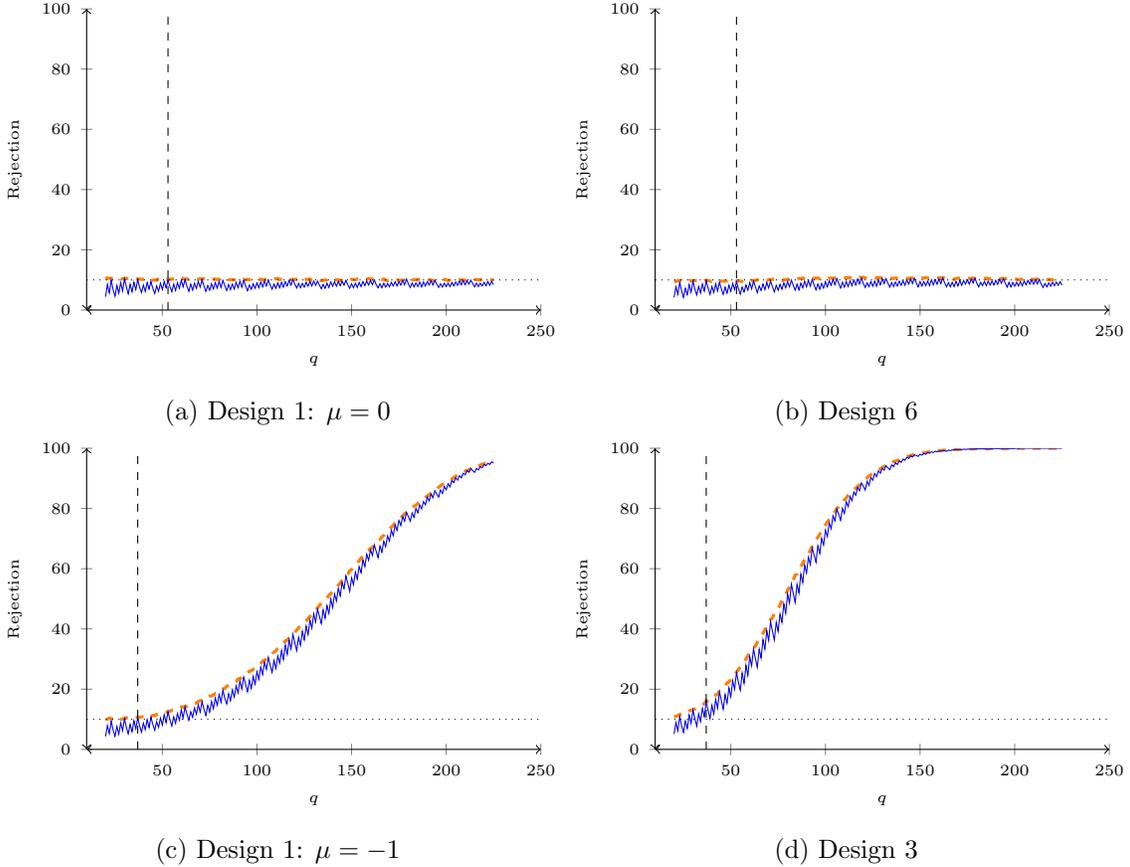
\begin{figure}[t!]
\centering
  %!TEX root = ../RDDcontinuity.tex
\pgfplotsset{every axis/.append style={
  axis x line=bottom,    % put the x axis in the middle
  axis y line=left,    % put the y axis in the middle
  axis line style={<->}, % arrows on the axis
  label style={font=\tiny},
  tick label style={font=\tiny}  
},
cycle list={
{draw=black, solid,dashed,very thick,orange},
{draw=black, solid,blue}, 
{draw=black,densely dashed,very thick,blue},
{draw=black, dashdotted, very thick,green},
{only marks, mark=asterisk}}}
\pgfplotstableread[col sep = comma]{dataplots.csv}\mydata

\begin{subfigure}[t]{0.45\textwidth}\centering  
    \begin{tikzpicture}[baseline]
      \begin{axis}[width=3.0in,height = 2.2in,xmin=10, xmax=250,ymin=0,ymax=100,xlabel=$q$,ylabel=Rejection]
      \addplot table[x index = {0}, y index = {1}] {\mydata};
      \addplot table[x index = {0}, y index = {2}] {\mydata};
      \draw[dashed] (53,0) -- (53,100); 
      \draw[dotted] (0,10) -- (250,10); 
      \end{axis}
    \end{tikzpicture}
    \caption{Design 1: $\mu=0$}\label{fig:q-d1}
\end{subfigure}
\begin{subfigure}[t]{0.45\textwidth}\centering  
    \begin{tikzpicture}[baseline]
      \begin{axis}[width=3.0in,height = 2.2in,xmin=10, xmax=250,ymin=0,ymax=100,xlabel=$q$,ylabel=Rejection]
      \addplot table[x index = {0}, y index = {3}] {\mydata};
      \addplot table[x index = {0}, y index = {4}] {\mydata};
      \draw[dashed] (53,0) -- (53,100); 
      \draw[dotted] (0,10) -- (250,10); 
      \end{axis}
    \end{tikzpicture}
    \caption{Design 6}\label{fig:q-d6}
\end{subfigure}

\begin{subfigure}[t]{0.45\textwidth}\centering  
    \begin{tikzpicture}[baseline]
      \begin{axis}[width=3.0in,height = 2.2in,xmin=10, xmax=250,ymin=0,ymax=100,xlabel=$q$,ylabel=Rejection]
      \addplot table[x index = {0}, y index = {5}] {\mydata};
      \addplot table[x index = {0}, y index = {6}] {\mydata};
      \draw[dashed] (37,0) -- (37,100); 
      \draw[dotted] (0,10) -- (250,10); 
X      \end{axis}
    \end{tikzpicture}
    \caption{Design 1: $\mu=-1$}\label{fig:q-d11}
\end{subfigure}
\begin{subfigure}[t]{0.45\textwidth}\centering  
    \begin{tikzpicture}[baseline]
      \begin{axis}[width=3.0in,height = 2.2in,xmin=10, xmax=250,ymin=0,ymax=100,xlabel=$q$,ylabel=Rejection]
      \addplot table[x index = {0}, y index = {7}] {\mydata};
      \addplot table[x index = {0}, y index = {8}] {\mydata};
      \draw[dashed] (37,0) -- (37,100); 
      \draw[dotted] (0,10) -- (250,10); 
      \end{axis}
    \end{tikzpicture}
    \caption{Design 3}\label{fig:q-d3}
\end{subfigure}
\caption{\footnotesize{Rejection probabilities of AS-NR (solid blue line) and AS-R (dashed orange line) as a function of $q$ for $n=1,000$. The vertical dashed line denotes the value of $\hat q_{\rm irot}$ and the horizontal dotted line the value of $\alpha$.}}
\label{fig:sensitivity.q}
\end{figure}

Tables \ref{tab:T1000} and \ref{tab:T5000} report rejection probabilities under the null and alternative hypotheses for the six designs we consider and for sample sizes of $n=1,000$ and $n=5,000$, respectively. We start by discussing the results under the null hypothesis. AS-NR delivers rejection probabilities under the null hypothesis closer to the nominal level than those delivered by McC and CJM in most of the designs. The two empirically motivated designs (Designs 3 and 6) illustrate this feature clearly. Designs 4 and 5 also show big differences in performance, both in cases where AS-NR delivers rejection rates equal to the nominal level (Design 5) and McC and CJM severely over-reject; as well as in cases where all tests over-reject (Design 4, $\kappa=0.05$) but AS-NR is relatively closer to the nominal level. A particularly difficult case for AS-NR is Design 1 with $\mu=-2$, where the probability of getting observations to the right of the cut-off is below $2\%$. This showcases the satisfactory performance of our data-dependent rule $\hat q_{\rm irot}$, which takes the lowest value in that particular design. Tables \ref{tab:T1000} and \ref{tab:T5000} also show negligible differences between the randomized (AS-R) and non-randomized (AS-NR) versions of our test, consistent with our discussion in Remark \ref{rem:non-randomized}.

To describe the performance of the different tests under the alternative hypothesis, we focus on designs where the rejection probability under the null hypothesis is close to the nominal level for all tests: Design 1, Design 2, Design 4 with $\kappa = 0.25$, and Design 6. In those cases, we see that AS-NR has competitive power, and can sometimes even be the test with the highest rejection probability under the alternative hypothesis. For $n=1,000$, AS-NR delivers the highest rejection probability under the alternative hypothesis in Design 1 for all values of $\mu$ and Design 4 with $\kappa=0.25$. In the rest of the cases under consideration, McC exhibits the highest power and is followed by AS-NR. The results for $n=5,000$ are qualitatively similar, with a few exceptions. McC has the highest rejection probability in Design 1 with $\mu=-1$, and CJM are have the second highest rejection probability in Design 2 with $\lambda = \frac{1}{3}.$  

Table \ref{tab:q-irot} shows the mean values of $\hat q_{\rm irot}$ across simulations for all designs and sample sizes. As described in Section \ref{sec:ourtest}, $\hat q_{\rm irot}$ takes into account both the slope and the magnitude of the density at the cut-off. As a result, $\hat q_{\rm irot}$ is relatively high in designs with flat density at the cut-off and high $f_Z(0)$ (e.g., Design 1 with $\mu=0$) and relatively low in designs with steep slopes or low $f_{Z}(0)$ (e.g., Design 1 with $\mu=-2$ or Design 2 with $\lambda=1$). Table \ref{tab:q-irot}  also reports the average number of observations in $[-h_{n,L},h_{n,R}]$ for McC and CJM, where $h_{n,L}$ and $h_{n,R}$ are the left and right bandwidths used to estimate $f^{-}_Z(0)$ and $f^{+}_Z(0)$, respectively (in the case of McC, $h_{n,L}=h_{n,R}$). In comparison, AS-NR uses significantly fewer observations than either McC or CJM, a feature that may support the the asymptotic framework in Section \ref{sec:fixed-q}. Finally, and to gain further insight on the sensitivity of our test to the choice of $q$, Figure \ref{fig:sensitivity.q} displays the rejection probabilities of AS-NR and AS-R as a function of $q$ in two types of designs. In the top row we illustrate two designs where the rejection probability is mostly insensitive to the choice of $q$ (Design 1 with $\mu=0$ and Design 6). These are designs where the density is rather flat around the cut-off so increasing $q$ does not deteriorate the performance of our test. In the bottom row we illustrate two designs where the rejection probability is highly sensitive to the choice of $q$ (Design 1 with $\mu=-1$ and Design 3). These are designs that feature a steep density at the cut-off (also low in Design 1) so increasing $q$ very quickly deteriorates the performance of the test under the null hypothesis. The data-dependent rule $\hat q_{\rm irot}$ is displayed in each case with a vertical dashed line and seems to be doing a good job at choosing relatively smaller values in the sensitive cases. 

We conclude this section with two final remarks. First, one could compare the results in Tables \ref{tab:T1000} and \ref{tab:T5000} for a fix value of $q$ to appreciate the results in Section \ref{sec:fixed-q}. For example, taking $q=75$, the rejection probability in Design 1 with $\mu=-2$ and Design 3 are $99.8$ and $40.1$, respectively, when $n=1,000$. The same numbers when $n=5,000$ are $26$ and $8.4$, respectively, which are closer to the nominal level as predicted by our results. Second, at the request of a referee, the results for $n=1,000,000$ and $\alpha=1\%$ are available upon request. Notably, AS-NR with the data-dependent rule $\hat q_{\rm irot}$ delivers rejection probabilities under $H_0$ equal to $\alpha$ across all designs when $n=1,000,000$ whereas McC and/or CJM\textbf{} still significantly over-reject for Designs 3, 5, and 6. 

\section{Empirical illustration}\label{sec:application}

In this section we briefly reevaluate the validity of the design in \cite{lee:08}. Lee studies the benefits of incumbency on electoral outcomes using a discontinuity constructed with the insight that the party with the majority wins. Specifically, the running variable $Z$ is the difference in vote shares between Democrats and Republicans in a house election; see Figure \ref{fig:Designs}(f) for a graphical illustration of the density of $Z$. The assignment rule then takes a cut-off value of zero that determines the treatment of incumbency to the Democratic candidate, which is used to study their outcome in the next election. The total number of observations is 6,559 with 2,740 below the cut-off. The dataset is publicly available at \url{http://economics.mit.edu/faculty/angrist/data1/mhe}. 

Lee assesses the credibility of the design in this application by inspecting discontinuities in means of the baseline covariates, but mentions in footnote 19 the possibility of using the test proposed by \cite{mccrary:08}. Here, we frame the validity of the design in terms of the hypothesis in \eqref{eq:null} and use the approximate sign test we describe in Section \ref{sec:ourtest}, using $\hat q_{\rm irot}$ as our default choice for the number of observations $q$. This test delivers a $p$-value of $0.55$ for $S_n = 73$ out of $\hat q_{\rm irot}=138$ observations. The null hypothesis of continuity of the density is therefore not rejected. 

\section{Concluding remarks}\label{sec:conclusion}
This paper presents an approximate sign test based on $g$-order statistics for testing the continuity of a density at a point in RDD. We study the properties of this test under two asymptotic frameworks; one in which the number $q$ of observations local to the cut-off is fixed as the sample size $n$ diverges to infinity, and one in which $q$ diverges to infinity slowly as $n$ diverges to infinity. We show that the test has limiting rejection probability under the null hypothesis not exceeding the nominal level in both asymptotic frameworks under similar and arguably mild conditions. More importantly, our test is easy to implement, asymptotically valid under weaker conditions than those used by competing methods, exhibits finite sample validity under stronger conditions than those needed for its asymptotic validity, and delivers competitive power in simulations. 

A final aspect we would like to highlight of our test is its simplicity. The test only requires to count the number of non-negative observations out of the $q$ observations closest to the cut-off (this is all that is required to compute the p-value in \eqref{eq:p-value}), and does not involve kernels, local polynomials, bias correction, or bandwidth choices. Importantly, we have developed the \verb rdcont  \verb+Stata+ package that allows for effortless implementation of the test we propose in this paper.

%-----------------------------------------------------------------------
%\newpage
\appendix
\setcounter{equation}{0}
\renewcommand{\theequation}{\Alph{section}-\arabic{equation}}
\small
%!TEX root = ../RDDcontinuity.tex
\section{Proof of the main results}\label{app:proofs}

\subsection{Proof of Theorem \ref{thm:g-order-stats}}

Throughout the proof we use $\{Z^*_{i}:1\le i\le n\} $ as defined in Lemma \ref{lem:Z-ast}, which in turn allow us to apply \citet[Theorem 1]{kaufmann/reiss:92} later in the proof, when invoking Lemma \ref{lemma:KR}.

Let $Z^{\ast}_{g,(1)},\dots,Z^{\ast}_{g,(q)}$ denote the $q$ values of $\{Z^{\ast}_i:1\le i\le n\}$ that are induced by the $q$ smallest values of $\{g(Z^{\ast}_i)=|Z^{\ast}_i|:1\le i\le n\}$ and let 
\begin{equation*}\label{eq:Agj-ast}
  A^{\ast}_{g,(j)} \equiv I\{Z^{\ast}_{g,(j)}\ge 0\} \text{ for } 1\le j\le q~
\end{equation*}
and 
\begin{equation}\label{eq:Sn-ast}
  S^{\ast}_n \equiv \sum_{j=1}^q A^{\ast}_{g,(j)}~.
\end{equation}

It is convenient to introduce the following notation. Let $p_{q}(s|\pi)$ denote the probability mass function (pmf) of ${\rm Bi}(q,\pi)$ with $\pi\in [0,1]$, i.e.,
\begin{equation}\label{eq:joint-pmf}
	p_{q}(s|\pi)= \binom{q}{s}\pi^s(1-\pi)^{q-s} ~.
\end{equation}
Note that $p_{q}(s|\pi)$ is continuous in $\pi$.

Next, consider $S_n$ in \eqref{eq:Sn}. Note that $S_n$ has support $\mathbf N_q\equiv \{0,1,\dots,q\}$ for all $n\in \mathbf N$, and so its CDF at any $x\in \mathbf R$ is $\sum_{s \in \mathbf N_q} P\{S_n=s\}I\{s\leq x\}$. From this, we conclude that $S_n \overset{d}{\to} S \sim {\rm Bi}(q,\pi_f)$ follows from showing that $P\{S_n=s\} \to p_{q}(s|\pi_f) $ for all $s\in \mathbf N_q$. To this end, consider the following derivation for arbitrary $s\in N_{q}$ and $\epsilon <\delta /2$ with $\delta $ as in Assumption \ref{ass:A}.
\begin{align}
|P\{S_{n}=s\}-p_{q}(s|\pi _{f})| &\leq R_{n,1}+|P\{S_{n}^{\ast }=s\}-p_{q}(s|\pi _{f})| \notag \\ & =R_{n,1}+\vert E[ P\{S_{n}^{\ast }=s~|~|Z_{g,(q+1)}^{\ast }|\}-p_{q}(s|\pi _{f})] \vert \notag \\
& \leq R_{n,1}+E\vert P\{S_{n}^{\ast }=s~|~|Z_{g,(q+1)}^{\ast }|\}-p_{q}(s|\pi _{f})\vert \notag \\
& =R_{n,1}+E[ \vert P\{S_{n}^{\ast }=s~|~|Z_{g,(q+1)}^{\ast }|\}-p_{q}(s|\pi _{f})\vert I\{|Z_{g,(q+1)}^{\ast }|\leq\epsilon \}] \notag \\
& \quad +E[ \vert P\{S_{n}^{\ast }=s~|~|Z_{g,(q+1)}^{\ast }|\}-p_{q}(s|\pi _{f})\vert I\{|Z_{g,(q+1)}^{\ast }|> \epsilon \} ] \notag \\
& \leq R_{n,1}+R_{2}( \varepsilon ) +R_{n,3}( \varepsilon ) ~,\label{eq:derivThm41}
\end{align}
where the first inequality follows from the triangle inequality and by setting $ R_{n,1}\equiv |P\{S_{n}=s\}-P\{S_{n}^{\ast }=s\}|$, the first equality follows from the law of iterated expectations, the second inequality follows from Jensen's inequality, and the last inequality follows from Lemma \ref{lemma:KR} and by setting $R_{2}( \varepsilon ) \equiv \sup_{r\leq\epsilon }\vert p_{q}(s|\pi (r))-p_{q}(s|\pi _{f})\vert $, and $ R_{n,3}( \varepsilon ) \equiv P\{|Z_{g,(q+1)}^{\ast }|>\epsilon \}$. By computing sequential limits $n\to \infty $ and $ \epsilon \downarrow 0$, we now show that right hand side of \eqref{eq:derivThm41} converges to zero. As $n\to \infty $, $ R_{n,1}=o( 1) $ by Lemma \ref{lem:almost-surely}(b) and $R_{n,3}( \varepsilon ) =o( 1) $ by Lemma \ref{lem:almost-surely}(a). By then taking $\epsilon \downarrow 0$, Lemma \ref{lem:Cutoff} implies that $\sup_{r\leq \epsilon }|\pi (r)-\pi _{f}|=o( 1) $, and this and the continuity of $p_{q}(s|\pi )$ in $\pi $ implies that $R_{2}( \varepsilon ) =o( 1) $. $\blacksquare$

% %-----------------------------------------------

\subsection{Proof of Theorem \ref{thm:main}}
By the definition of $\phi(S_n)$ in \eqref{eq:testmain} and the expressions of $T(S_n)$ in \eqref{eq:T} and $c_q(\alpha)$ in \eqref{eq:cq}, 
\begin{equation*}
	E[\phi(S_n)] = P\{S_n<b_q(\alpha)\} + P\{S_n>q-b_q(\alpha)\} + a_q(\alpha)\left(P\{S_n=b_q(\alpha)\}+P\{S_n=q-b_q(\alpha)\}\right)~. 
\end{equation*}
Theorem \ref{thm:g-order-stats} shows that $P\{S_n=s\} = P\{S=s\} +o(1)$ for all $s\in \mathbf N_q\equiv \{0,1,\dots,q\}$, where $S\sim {\rm Bi}(q,\pi_f)$ and $\pi_f$ is as in Theorem \ref{thm:g-order-stats}. It follows from this result and the above display that $E[\phi(S_n)]\to E[\phi(S)]$ as $n\to\infty$, where 
\begin{equation}\label{eq:EphiS}
	E[\phi(S)] = P\{S<b_q(\alpha)\} + P\{S>q-b_q(\alpha)\} + a_q(\alpha)\left(P\{S=b_q(\alpha)\}+P\{S=q-b_q(\alpha)\}\right)~. 
\end{equation}
We complete the proof by analyzing \eqref{eq:EphiS} under $H_0$ and $H_1$ in \eqref{eq:null}. 

Under $H_0$ in \eqref{eq:null}, $S\sim {\rm Bi}(q,\frac{1}{2})$. In this case,
\begin{equation*}
	P\{S<b_q(\alpha)\} + P\{S>q-b_q(\alpha)\}=2\Psi_q(b_q(\alpha)-1)~,
\end{equation*}
where we used that $b_q(\alpha)\in\{0,1,\dots,\lfloor \frac{q}{2}\rfloor \}$ and $P\{S<b\}=P\{S>q-b\}$ for any $b\in\{0,\dots,\lfloor \frac{q}{2}\rfloor\}$ when $\pi_f=\frac{1}{2}$. In addition, 
\begin{equation*}\label{eq:rej-equal}
	a_q(\alpha)\left(P\{S=b_q(\alpha)\}+P\{S=q-b_q(\alpha)\}\right) = 2a_q(\alpha)\frac{1}{2^{q}}\binom{q}{b_q(\alpha)}~,
\end{equation*}
where we used that $\binom{q}{C}=\binom{q}{q-C}$ for any $C\in\{0,\dots,q\}$. Therefore,
\begin{equation}\label{eq:limit-size}
	E[\phi(S)] = 2\Psi_q(b_q(\alpha)-1)+\frac{a_q(\alpha)}{2^{q-1}}\binom{q}{b_q(\alpha)}=\alpha~,
\end{equation}
where the last equality follows by definition of $a_q(\alpha)$.

By \citet[][Example 4.2.1]{lehmann/romano:05} (with $p_{0}=1/2$), $\phi (\cdot)$ in \eqref{eq:testmain} is an unbiased test for \eqref{eq:null}. From this and \eqref{eq:limit-size}, it follows that $E [ \phi ( S)] \geq \alpha $ under $H_{1}$ in \eqref{eq:null}, as desired. $\blacksquare$

 % %-----------------------------------------------

\subsection{Proof of Theorem \ref{thm:largeq-size}}
For $S_n$ as in \eqref{eq:Sn} and $S_n^*$ as in \eqref{eq:Sn-ast}, let 
\begin{equation*}
\xi_q(\pi)\equiv \sqrt{q}\left(\frac{1}{q}S_n -\pi \right) \quad \text{ and }\quad \xi_q^{\ast}(\pi)\equiv\sqrt{q}\left(\frac{1}{q}S^*_n -\pi \right)~.
\end{equation*}
For any $\pi\in (0,1)$ and for $S\sim {\rm Bi}(q,\pi)$, let $\Psi_q(x|\pi)$ denote the CDF of $S$ and let
\begin{equation*}
	J_q(x|\pi) \equiv P\left\lbrace \sqrt{q}\left(\frac{1}{q}S -\pi \right)\le x \right\rbrace~.
\end{equation*}

It suffices to show that for any $\eta>0$, there exists $N$ such that $\forall n\ge N$,
\begin{equation*}
	\left|P\{\xi_q(\pi_f)\le x\}-\Phi\left(\frac{x}{\sqrt{\pi_f(1-\pi_f)}}\right) \right|\le \eta~.
\end{equation*}
Let $\{\epsilon_q:q\geq 1 \}$ be a sequence in $(0,\delta/2)$ that satisfies $\sqrt{q}\epsilon_q\to 0$ and $\epsilon_q\frac{n}{q}\to \infty$. Since $\frac{q^{3/2}}{n}\to 0$ these conditions occur for all $q$ sufficiently large if we set $\epsilon_q =\frac{1}{q^{1/2}}\left(\log\left(\frac{n}{q^{3/2}}\right)\right)^{-1}$. Consider the following decomposition for $x\in \mathbf R$, 
\begin{equation}\label{eq:partition-bar0}
	P\{\xi_q(\pi_f)\le x\} = \bar{R}_{n,1} + \bar{R}_{n,2} + \bar{R}_{n,3}~,
\end{equation}
with 
\begin{align*}
	\bar{R}_{n,1} & ~\equiv~ P\{\xi_q(\pi_f)\le x\}-P\{\xi_q^{\ast}(\pi_f)\le x\}   \\
	\bar{R}_{n,2} & ~\equiv~ E[ P\{\xi_q^{\ast}(\pi_f)\le x~|~|Z^{\ast}_{g,(q+1)}|\}I\{|Z^{\ast}_{g,(q+1)}|>{\epsilon_q}\} ]  \\
	\bar{R}_{n,3} & ~\equiv~ E[ P\{\xi_q^{\ast}(\pi_f)\le x~|~|Z^{\ast}_{g,(q+1)}|\}I\{|Z^{\ast}_{g,(q+1)}|\leq{\epsilon_q}\} ]~.
\end{align*}
First, Lemma \ref{lem:almost-surely}(b) implies that $\bar{R}_{n,1} = o(1)$. Second, $\bar{R}_{n,2}=o(1)$ follows from 
\begin{equation*}
	0\le \bar{R}_{n,2} \le P\{|Z^{\ast}_{g,(q+1)}|> \epsilon_q\}=P\Big\{\frac{n}{q}|Z^{\ast}_{g,(q+1)}|> \epsilon_q\frac{n}{q}\Big\}=o(1)~,
\end{equation*}
where the last equality follows from Lemma \ref{lem:rate-Zg} and $\epsilon_q\frac{n}{q}\to \infty$. Finally, let $\pi^{+}_q \equiv \pi_f + \frac{1}{2}\frac{\epsilon_q C_P}{f_{Z}^{+}(0)+f_{Z}^{-}(0)}$ and consider the following derivation,
\begin{align}
	\bar{R}_{n,3} 
	& \ge P\{|Z^{\ast}_{g,(q+1)}|\le \epsilon_q\}\inf_{r\le \epsilon_q}P\left\lbrace S_n^{\ast}\le \sqrt{q}x+q\pi_f~\Big|~|Z^{\ast}_{g,(q+1)}|=r\right\rbrace \notag\\
	& = P\{|Z^{\ast}_{g,(q+1)}|\le \epsilon_q\}\inf_{r\le \epsilon_q} \Psi_q\left(\sqrt{q}x+q\pi_f\Big |\pi(r)\right)\notag\\
	& \ge P\{|Z^{\ast}_{g,(q+1)}|\le \epsilon_q\}\Psi_q\left(\sqrt{q}x+q\pi_f\Big |\pi^{+}_q\right)\notag\\
	& = P\{|Z^{\ast}_{g,(q+1)}|\le \epsilon_q\}J_q\left(x-\sqrt{q}(\pi_q^{+}-\pi_f)\Big |\pi^{+}_q\right)\notag\\
	&=P\{|Z^{\ast}_{g,(q+1)}|\le \epsilon_q\}J_q\left(x-\frac{1}{2}\frac{\sqrt{q}\epsilon_q C_P}{f_{Z}^{+}(0)+f_{Z}^{-}(0)} \Bigg|\pi^{+}_q\right),\label{eq:R3_1}
\end{align}
where the first equality uses $\pi(r)=P\{Z>0~|~|Z|<r\}$ and follows from Lemma \ref{lemma:KR}, the second inequality follows from $\Psi_q(x|\pi)$ being decreasing in $\pi$ and $\pi(r)\le \pi^{+}_q$ for $r\le \epsilon_q$ from Lemma \ref{lem:bound_on_pi}, and the last equality follows from the definition of $\pi^{+}_q$. By an analogous argument,
\begin{align}
	\bar{R}_{n,3} &\le P\{|Z^{\ast}_{g,(q+1)}|\le \epsilon_q\}\sup_{r\le \epsilon_q}P\left\lbrace S_n^{\ast}\le \sqrt{q}x+q\pi_f~\Big|~|Z^{\ast}_{g,(q+1)}|=r\right\rbrace \notag\\
	& \le P\{|Z^{\ast}_{g,(q+1)}|\le \epsilon_q\}\Psi_q\left(\sqrt{q}x+q\pi_f\Big |\pi^{-}_q\right)\notag\\
	&\le P\{|Z^{\ast}_{g,(q+1)}|\le \epsilon_q\}J_q\left(x+\frac{1}{2}\frac{\sqrt{q}\epsilon_qC_P}{f_{Z}^{+}(0)+f_{Z}^{-}(0)}\Bigg|\pi^{-}_q\right), \label{eq:R3_2}
\end{align}
where in this case we define $\pi^{-}_q \equiv \pi_f - \frac{1}{2}\frac{\epsilon_q C_P}{f_{Z}^{+}(0)+f_{Z}^{-}(0)}$. To complete the proof, it suffices to show that that the right-hand side expressions of \eqref{eq:R3_1} and \eqref{eq:R3_2} converge to $\Phi\left(\frac{x}{\sqrt{\pi_f(1-\pi_f)}}\right)$. We only show the result for \eqref{eq:R3_1}, as the result for \eqref{eq:R3_2} is analogous.

It follows by the Berry-Esseen theorem that 
\begin{equation}
	\left|J_q\left(x-\frac{1}{2}\frac{\sqrt{q}\epsilon_q C_P}{f_{Z}^{+}(0)+f_{Z}^{-}(0)}\Bigg |\pi^{+}_q\right)- \Phi\left(\frac{x-\frac{1}{2}\frac{\sqrt{q}\epsilon_q C_P}{f_{Z}^{+}(0)+f_{Z}^{-}(0)}}{\sqrt{\pi^{+}_q(1-\pi^{+}_q)}}\right) \right|\le \frac{1}{2\sqrt{q}}\frac{((\pi^{+}_q)^2+(1-\pi^{+}_q)^2)}{\sqrt{\pi^{+}_q(1-\pi^{+}_q)}}\to0~,\label{eq:R3_3}
\end{equation}
where the convergence follows from $q\to \infty $ and $\pi _{q}^{+}\to \pi _{f}\in (0,1)$. Since $\sqrt{q}\epsilon _{q}\to 0$, the continuity of the standard normal CDF implies that
\begin{equation}
	\left| \Phi\left(\frac{x-\frac{1}{2}\frac{\sqrt{q}\epsilon_q C_P}{f_{Z}^{+}(0)+f_{Z}^{-}(0)}}{\sqrt{\pi^{+}_q(1-\pi^{+}_q)}}\right)-\Phi\left(\frac{x}{\sqrt{\pi_f(1-\pi_f)}}\right)\right| \to 0~.\label{eq:R3_4}
\end{equation}
Finally, Lemma \ref{lem:rate-Zg} and $\epsilon_q\frac{n}{q}\to \infty$ imply that 
\begin{equation}
	P\{|Z^{\ast}_{g,(q+1)}|\le \epsilon_q\}=P\Big\{\frac{n}{q}|Z^{\ast}_{g,(q+1)}|\le \frac{n}{q}\epsilon_q\Big\}\to 1~. \label{eq:R3_5}
\end{equation}
By combining \eqref{eq:R3_3}, \eqref{eq:R3_4}, and \eqref{eq:R3_5}, \eqref{eq:main-thm4.3} follows. 

We now conclude the proof by showing parts (a)-(c) of the theorem. By the definition of $c_q(\alpha)$ in \eqref{eq:cq} and the central limit theorem, it follows that for any $\alpha\in(0,1)$,
\begin{equation}
 	c_q(\alpha)\to \frac{1}{2}z_{\alpha/2}
 \end{equation} 
as $q\to \infty$, where $z_{\alpha/2}$ is the $(1-\frac{\alpha}{2})$-quantile of $N(0,1)$. Since $q\to \infty$ as $n\to \infty$, this implies that $c_q(\alpha)$ converges to the $(1-\alpha/2)$-quantile of $N(0,1/4)$ as $n\to \infty$. Next, notice that 
\begin{equation}
  \sqrt{q}\left(\frac{1}{q}S_n -\frac{1}{2} \right)~=~\sqrt{q}\left(\frac{1}{q}S_n -\pi_f \right) +\sqrt{q}\left(\pi_f-\frac{1}{2} \right)~.
  \label{eq:aux_convergence}
\end{equation}
Under $H_0$ in \eqref{eq:null}, $\pi_f=\frac{1}{2}$, and so \eqref{eq:main-thm4.3} implies that the expression in \eqref{eq:aux_convergence} converges in distribution to $N(0,1/4)$ as $n\to \infty$. From here, part (a) follows. Parts (b) and (c) follow from analogous arguments based on \eqref{eq:main-thm4.3}, \eqref{eq:aux_convergence}, and the convergence of $c_q(\alpha)$ to the $(1-\alpha/2)$-quantile of $N(0,1/4)$ as $n\to \infty$. $\blacksquare$
%!TEX root = ../RDDcontinuity.tex
\section{Auxiliary Results}\label{app:lemmas}

\begin{lemma}\label{lem:Z-ast}%this lemma was {lem:Translation}
	Let $\delta>0$ be as in Assumption \ref{ass:A} and let $\{\upsilon_{i}:1\le i\le n\} $ be an i.i.d.\ sample such that $\upsilon_{i}\sim U(-\frac{\delta}{2},\frac{\delta}{2})$ independent of $Z^{(n)}$. Define the sequence of i.i.d.\ random variables $\{Z^*_{i}:1\le i\le n\} $ as
	\begin{equation*}
		Z_{i}^{\ast}\equiv Z_{i}+\upsilon_{i}I\{|Z_{i}|\ge \delta\}~.
	\end{equation*}
	Then,
	\begin{enumerate}[(a)]
	 	\item The distribution function of $|Z^*|$ is continuous on $\mathbf R$. 
	 	\item For any $r\in (0,\frac{\delta}{2})$, 
		\begin{equation}\label{eq:same-prob}
			P\{Z^{\ast }\geq 0~|~\vert Z^{\ast }\vert <r\} =P\{Z \geq 0~|~ |Z| <r\}~.
		\end{equation}	
		\item For any $r>0$, $P\{|Z^*|<r\}>0$. 
	\end{enumerate} 
\end{lemma}

\begin{proof}
	To prove part (a), let $E\subset \mathbf R$ be a set of zero Lebesgue measure and note that  
	\begin{eqnarray*}
		P\{|Z^{\ast }|\in E\} &=& P\{|Z+\upsilon I\{|Z|\ge\delta\}|\in E\} \\
		&=& P\{|Z+\upsilon I\{|Z|\ge\delta\}|\in E \cap |Z|\ge\delta\} 
			+ P\{|Z+\upsilon I\{|Z|\ge\delta\}|\in E \cap |Z|< \delta\} \\
		&=& P\{|Z+\upsilon|\in E \cap |Z|\ge\delta\} + P\{|Z|\in E \cap |Z|< \delta\} \\
		&\leq& P\{|Z+\upsilon|\in E\}+ P\{|Z|\in E \cap (0,\delta)\}=0~,
	\end{eqnarray*}
	where the last equality holds because the distribution function of $|Z+\upsilon |$ is continuous and $E\cap(0,\delta)$ is a subset of zero Lebesgue measure in the set where the distribution function of $|Z|$ is assumed to be continuous.

	For part (b), note that for any $r\in (0,\frac{\delta}{2}) $, $\vert Z^{\ast }\vert <r$ implies that $Z=Z^{\ast }$ and \eqref{eq:same-prob} follows. 

	For part (c), use again that $P\{|Z^{\ast }|<r\}=P\{|Z|<r\}$ whenever $r\in (0,\frac{\delta}{2})$. By Assumption \ref{ass:A}, for any $0<\epsilon <\delta$, 
	\begin{equation*}
	\frac{1}{\epsilon}P \{|Z|<\epsilon\}=\frac{1}{\epsilon} \int_{-\epsilon} ^{\epsilon} f_{Z}(z) dz~.
	\end{equation*}
	Taking limits as $\epsilon\downarrow 0$, using L'H\^{o}pital's rule, and invoking Assumption \ref{ass:A}(ii) shows that $\lim_{\epsilon\downarrow 0}\frac{1}{\epsilon}P \{|Z|<\epsilon\}=f_Z^{+}(0)+f_Z^{-}(0)>0$. Thus, there exists $\bar{\epsilon}<\delta$ such that $P \{|Z|< \epsilon \}>0$ for all $\epsilon \in (0,\bar{\epsilon})$ and so $P\{|Z|<r\}>0$ for all $r\in\mathbf R$. This completes the proof. 
\end{proof}

\begin{lemma}\label{lem:Cutoff}
	Let Assumptions \ref{ass:A}(i) and \ref{ass:A}(ii) hold and let $\pi_f$ be defined as in Theorem \ref{thm:g-order-stats}. Then, for all $\mu>0$, there exists $\epsilon >0$ s.t. 
	\begin{equation*}
	\sup_{r\leq \epsilon }\left|P\{Z\geq 0~|~|Z|<r\}-\pi_f\right| \leq \mu~.
	\end{equation*}
\end{lemma}

\begin{proof}
	First note that, under Assumption \ref{ass:A}(ii), the proof of Lemma \ref{lem:Z-ast} shows that $P\{|Z|<r\}>0$ for all $r\in\mathbf R$. It follows that 
	\begin{align*}
		P\{Z\geq 0~|~|Z|<\epsilon\}~=~ \frac{\frac{1}{\epsilon} \int_{0} ^{\epsilon} f_{Z}(z) dz}{\frac{1}{\epsilon} \int_{-\epsilon} ^{\epsilon} f_{Z}(z) dz} 
		~=~\frac{f^{+}_Z(0)}{f^{+}_{Z}(0)+f^{-}_{Z}(0) } + \Delta_{\epsilon}~,
	\end{align*}
	where the last equality holds for $\Delta_{\epsilon}\to 0$ as $\epsilon\to 0$ by using L'H\^{o}pital's rule and Assumption \ref{ass:A}(ii). The result then follows by definition of $\pi_f$.
\end{proof}

\begin{lemma}\label{lem:almost-surely}
	Let Assumptions \ref{ass:A}(i) and \ref{ass:A}(ii) hold and $\frac{q}{n}\to 0$ as $n\to\infty$. Then, 
	\begin{enumerate}[(a)]
		\item For any $\epsilon \in (0,\frac{\delta}{2})$, $P\{\liminf_{n\to\infty} \{|Z^{\ast}_{g,(q+1)}|\leq \epsilon \}\}=P\{\liminf_{n\to\infty} \{|Z_{g,(q+1)}|\leq \epsilon \}\} =1$.
		\item $P\{\liminf_{n\to\infty} \{S_n=S_n^{\ast}\} \}=1$, where $S_n$ is as in \eqref{eq:Sn} and $S_n^{\ast}$ is as in \eqref{eq:Sn-ast}.
	\end{enumerate}
\end{lemma}

\begin{proof}
	Fix $\epsilon \in (0,\frac{\delta}{2})$ arbitrarily and set $N_{n}\equiv \sum_{i=1}^{n}I\{|Z_{i}|\leq \epsilon\}$. Note that $N_{n}\geq q+1$ implies that $Z^{\ast}_{i}=Z_{i}$ and $Z^{\ast}_{g,(j)}=Z_{g,(j)}$ for at least $q+1$ observations that are within an $\epsilon$-neighborhood of zero. It follows that for all these observations, $A_{g,(j)}=A^{\ast}_{g,(j)}$, $Z^{\ast}_{g,(j)}\le \epsilon$, and $Z_{g,(j)}\le \epsilon$. We conclude that $N_n\ge q+1$ implies 
	\begin{equation*}
		S_n = S_n^{\ast}, \quad Z^{\ast}_{g,(q+1)}\le \epsilon,\quad \text{and}\quad Z_{g,(q+1)}\le \epsilon ~.
	\end{equation*}
	Parts (a)-(b) thus follow from proving that $P\{\liminf_{n\to\infty}\{N_{n}\geq q+1\}\}=1$. To show this, note that $N_{n}\sim {\rm Bi}(n,P\{|Z|\le \epsilon\})$. Now set $\mu \equiv \frac{1}{2}P\{|Z|\le \epsilon\}$, which is positive by the proof of Lemma \ref{lem:Z-ast}. It follows that 
	\begin{align*}
		P\{\liminf_{n\to\infty} \{N_{n}\geq q+1\}\} 
		= P\left\lbrace \liminf_{n\to\infty} \left\lbrace \frac{N_{n}}{n}\geq \frac{q+1}{n}\right\rbrace\right\rbrace  
		\geq P\left\lbrace \liminf_{n\to\infty} \left\lbrace\frac{N_{n}}{n}\geq \mu \right\rbrace\right\rbrace=1~,
	\end{align*}
	where the inequality holds for all $n>(q+1)/\mu$, and the last equality follows by the strong law of large numbers, i.e., $N_{n}/n\overset{a.s.}{\to}2\mu >0$. 
\end{proof}

\begin{lemma} \label{lemma:KR} 
	Let Assumptions \ref{ass:A}(i) and \ref{ass:A}(ii) hold. Fix $r\in (0,\frac{\delta}{2})$ and $q\in \{1,\ldots ,n-1\}$ arbitrarily. Then, for all $s\in \mathbf N_q \equiv \{0,1,\dots,q\}$,
	\begin{equation*}
		P\{S_n^{\ast}=s~|~|Z^{\ast}_{g,(q+1)}|=r\}=p_{q}(s|\pi(r))
	\end{equation*}
	where $p_{q}(s|\pi(r))$ is the pmf defined in \eqref{eq:joint-pmf} with $\pi(r)\equiv P\{Z\geq 0~|~|Z|<r\}$.
\end{lemma}

\begin{proof}
	Let $X\equiv (|Z^*|,A^*)$ with $A^*=I\{Z^*\ge 0\}$ and note that the $g$-order statistics we defined in Section \ref{sec:ourtest} using $g=|\cdot|$, could be alternatively obtained using $X$ and $\tilde{g}$-order statistics where $\tilde{g}$ is now the projection into the first component of $X$, i.e.
	\begin{equation*}
		\tilde{g}(X) = |Z^*|~.
	\end{equation*}
	In this way, and for this particular choice of $\tilde{g}$, $\tilde{g}$-order statistics on $X$ deliver 
	\begin{equation*}
		X_{\tilde{g},(1)}\equiv (|Z^*|_{(1)},A^*_{[1]})\le_{\tilde{g}} (|Z^*|_{(2)},A^*_{[2]})\le_{\tilde{g}} \cdots \le_{\tilde{g}} (|Z^*|_{(n)},A^*_{[n]})\equiv X_{\tilde{g},(n)}~,
	\end{equation*}
	where the random variables $(A^*_{[1]},\dots,A^*_{[n]})$ are called \emph{induced order statistics} or \emph{concomitants} of order statistics, see \cite{david/galambos:74,bhattacharya:74}. 

	Let $\tilde{X}_1,\dots,\tilde{X}_q$ be i.i.d.\ bivariate random variables such that $\tilde{X} \overset{d}{=}\{X~|~\tilde{g}(X)<r\}$. Theorem 1 in \cite{kaufmann/reiss:92} states that 
	\begin{equation}\label{eq:KR-thm1}
		\{(X_{\tilde{g},(1)},\dots,X_{\tilde{g},(q)})~|~\tilde{g}(X_{\tilde g,(q+1)})=r\} \overset{d}{=} \{\tilde{X}_{\tilde{g},(1)},\dots,\tilde{X}_{\tilde{g},(q)}\}~,
	\end{equation}
	with $\tilde{X}_{\tilde{g},(1)},\dots,\tilde{X}_{\tilde{g},(q)}$ being the $\tilde{g}$-order statistics of $\tilde{X}_1,\dots,\tilde{X}_q$, provided that (i) $\tilde{g}(X)$ has a continuous distribution and (ii) $P\{\tilde{g}(X)<r\}>0$. Since $\tilde{g}(X)=|Z^{\ast}|$ has a continuous distribution by Lemma \ref{lem:Z-ast}(a) and $P\{\tilde{g}(X)<r\}=P\{|Z^{\ast}|<r\}>0$ by Lemma \ref{lem:Z-ast}(c), we use \eqref{eq:KR-thm1} to prove our result. 

	Next, note that we can re-write $S_n^{\ast}$ in \eqref{eq:Sn-ast} as a function of $(X_{\tilde{g},(1)},\dots,X_{\tilde{g},(q)})$ by using the function $h$ that projects into the second component of $X$, i.e.
	\begin{equation*}
	 	S_n^{\ast} = \sum_{j=1}^q A^{\ast}_{g,(j)}= \sum_{j=1}^q A^{\ast}_{[j]} = \sum_{j=1}^q h(X_{\tilde{g},(j)})~,
	 \end{equation*} 
	 where in the second equality we used that $A^{\ast}_{g,(j)}=A^{\ast}_{[j]}$ by definition. Using this characterization, it follows that
	 \begin{align*}
	 	P\{S_n^{\ast} =s~|~|Z^{\ast}_{g,(q+1)}|=r\} 
	 	&= P\Big\{\sum_{j=1}^q h(X_{\tilde{g},(j)})=s~|~|\tilde{g}(X_{\tilde g,(q+1)})|=r\Big\} \\
	 	&= P\Big\{\sum_{j=1}^q h(\tilde{X}_{\tilde{g},(j)})=s \Big\}\\
	 	&= P\Big\{\sum_{j=1}^q h(\tilde{X}_j)=s \Big\}\\
	 	& = p_{q}(s|\pi(r))~,
 	 \end{align*}
where the second equality follows from \eqref{eq:KR-thm1}, the third equality follows from $\sum_{j=1}^q h(\tilde{X}_{\tilde{g},(j)})=\sum_{j=1}^q h(\tilde{X}_j)$, and the last equality follows from $h(\tilde{X}_1),\dots,h(\tilde{X}_q)$ being i.i.d.\ bivariate random variables such that $h(\tilde{X}) \overset{d}{=}\{h(X)~|~\tilde{g}(X)<r\}$ and $\{h(X)~|~\tilde{g}(X)<r\}=\{I\{Z^{\ast}\ge 0\}~|~|Z^{\ast}|<r\}$ being distributed Bernoulli with parameter $\pi(r)= P\{Z^{\ast}\ge 0 ~|~|Z^{\ast}|<r\}$. Since $ P\{Z^{\ast}\ge 0 ~|~|Z^{\ast}|<r\}= P\{Z \ge 0 ~|~|Z|<r\}$ for $r\in (0,\frac{\delta}{2})$ by Lemma \ref{lem:Z-ast}(b), this completes the proof.  
	% NOTE: this is a finite sample result that applies to all $n\in \mathbb{N}$. 
\end{proof}

\begin{lemma}\label{lem:rate-Zg}
  Let Assumptions \ref{ass:A}(i') and \ref{ass:A}(ii) hold and suppose $q \to \infty$ and $\frac{q}{n} \to 0$ as $n\to \infty$. Then $$\frac{n}{q}|Z^{\ast}_{g,(q)}|~\overset{P}{\to}~\frac{1}{f^{+}_{Z}(0)+f^{-}_{Z}(0)}~.$$
\end{lemma}

\begin{proof}
For any $\epsilon>0$, it suffices to show that
\begin{align} 
P\Big\{\frac{n}{q}|Z^{\ast}_{g,(q)}|>\frac{1}{f^{+}_{Z}(0)+f^{-}_{Z}(0)}+\epsilon\Big\}\to 0~~~\text{and}~~~
P\Big\{\frac{n}{q}|Z^{\ast}_{g,(q)}|<\frac{1}{f^{+}_{Z}(0)+f^{-}_{Z}(0)}-\epsilon\Big\}\to 0~.\label{eq:conv_zero}
\end{align}
We only show the first result in \eqref{eq:conv_zero}, as the second one follows from  symmetric arguments. By definition, $|Z^{\ast}_{g,(q)}|=|Z^{\ast}|_{(q)}$ where $|Z^{\ast}|_{(q)}$ denotes the $q$-th order statistic of the absolute value of $Z^{\ast}$. Denote by $Q$ the CDF of $|Z^{\ast}|$ and by $U_{(q)}$ the $q$-th order statistic of a $U[0,1]$ distributed random variable. Lemma \ref{lem:Z-ast}(a) implies that $Q(\cdot)$ is a continuous CDF. Then, for $\bar{M}=\frac{1}{f^{+}_{Z}(0)+f^{-}_{Z}(0)}+\epsilon$, note that
\begin{align}
	P\Big\{ \frac{n}{q}|Z^{\ast}_{g,(q)}|>\bar{M}\Big\} &= P\Big\{ |Z^{\ast}|_{(q)}>\frac{q}{n}\bar{M}\Big\}\notag \\
	& = P\Big\{ U_{(q)}>Q\left(\frac{q}{n}\bar{M}\right)\Big\}\notag \\
	& = P\Big\{ \frac{n^{1/2}}{e_n}(U_{(q)}-\mu_n)>\frac{n^{1/2}}{e_n}\left(Q\left(\frac{q}{n}\bar{M}\right)-\mu_n\right)\Big\}\label{eq:uniforms}~,
\end{align}
where $e_n^2=\mu_n(1-\mu_n)$ and $\mu_n = q/(n+1)$. Letting $$ \gamma_n = e_n n^{1/2}\frac{1}{e^2_n}\left(Q\left(\frac{q}{n}\bar{M}\right)-\mu_n\right)~,$$ it follows from \eqref{eq:uniforms} and \citet[][Eq. (3.1.2) in Lemma 3.1.1]{reiss:89}  that $$ P\Big\{ \frac{n}{q}|Z^{\ast}_{g,(q)}|>\bar{M}\Big\}\le \exp\left( -\frac{\gamma_n^2}{3(1+\gamma_n/(e_n n^{1/2}))}\right)~.$$
To complete the proof it suffices to show that the right-hand side expression in the display above converges to zero. To this end, it suffices to show that $\gamma_n \to \infty$ and $\gamma_n/(e_n n^{1/2})$ converges to a positive constant. In turn, this follows from showing that
\begin{equation}
e_n n^{1/2}\to \infty~~~\text{and}~~~\frac{1}{e_n^2}\left(Q\left(\frac{q}{n}\bar{M}\right)-\mu_n\right)\to \epsilon\left(f^{+}_Z(0)+f^{-}_Z(0)\right)>0~,
\label{eq:to_show_above}
\end{equation}
where the limit of the second expression is positive by Assumption \ref{ass:A}(ii). To show the first result in \eqref{eq:to_show_above}, note that $\frac{q}{n} \to 0$ implies $\frac{(e_n n^{1/2})^2}{q} = \frac{\mu_n (1-\mu_n)}{q/n} = \frac{n}{n+1} (1-\frac{q}{n+1}) \to 1$. Combined with $q\to \infty$, this then implies that $e_n n^{1/2}\to \infty$. To show the second result in \eqref{eq:to_show_above}, note that $\frac{q}{n} \to 0$ implies that for $n$ sufficiently large we obtain $\frac{q}{n}\bar{M}<\delta$ and so $Q\left(\frac{q}{n}\bar{M}\right)=\int_{-\frac{q}{n}\bar{M}}^{\frac{q}{n}\bar{M}}f_Z(z)dz$ since $Z^{\ast}=Z$ on $(-\delta,\delta)$. Then,  
\begin{align*}
	\frac{1}{e_n^2}\left(Q\left(\frac{q}{n}\bar{M}\right)-\mu_n\right) 
	&= \frac{q/n}{\mu_n(1-\mu_n)}\left( \frac{n}{q}\int_{-\frac{q}{n}\bar{M}}^{\frac{q}{n}\bar{M}}f_Z(z)dz-\frac{n}{q}\mu_n\right)\\
	& = \frac{q/n}{\mu_n(1-\mu_n)}\left( \frac{n}{q}\int_0^{\frac{q}{n}\bar{M}}(f_Z(z)-f^{+}_Z(0))dz+\frac{n}{q}\int_{-\frac{q}{n}\bar{M}}^0(f_Z(z)-f^{-}_Z(0))dz\right)\\
	& + \frac{q/n}{\mu_n(1-\mu_n)}\left(  \frac{n}{q}\int_0^{\frac{q}{n}\bar{M}}f^{+}_Z(0)dz+ \frac{n}{q}\int_{-\frac{q}{n}\bar{M}}^0f^{-}_Z(0)dz-\frac{n}{q}\mu_n\right)\\
	&\to \bar{M}\left(f^{+}_Z(0)+f^{-}_Z(0)\right)-1=\epsilon\left(f^{+}_Z(0)+f^{-}_Z(0)\right)~,
\end{align*}
where the convergence follows from Assumption \ref{ass:A}(i') and $\frac{q}{n} \to 0$, which imply that$$ \Big|\frac{n}{q}\int_0^{\frac{q}{n}\bar{M}}(f_Z(z)-f^{+}_Z(0))dz\Big|\le C_P\bar{M}^2\frac{q}{n}\to 0 \quad \text{ and }\quad \Big|\frac{n}{q}\int_{-\frac{q}{n}\bar{M}}^0(f_Z(z)-f^{-}_Z(0))dz\Big|\le C_P\bar{M}^2\frac{q}{n}\to 0~,$$
and $\frac{q/n}{\mu_n(1-\mu_n)}\to 1$.
\end{proof}

\begin{lemma}\label{lem:bound_on_pi}
  Let Assumption \ref{ass:A}(i') hold, $\pi_f$ be as in Theorem \ref{thm:g-order-stats}, and $\pi(r) = P\{Z>0~|~|Z|< r\}$. Then, for any $r\in (0,\delta)$, $$ \left|\pi(r)-\pi_f \right|\le\frac{r}{2} \frac{C_P}{f^{+}_Z(0)+f^{-}_Z(0)}~ .$$
\end{lemma}

\begin{proof}
Fix $r \in (0,\delta)$ arbitrarily. Start by re-writing $\pi(r)$ as follows, 
\begin{align}
    \pi(r) &=\frac{\frac{1}{r}\int_{0}^{r}f_{Z}\left( z\right) dz}{\frac{1}{r}\int_{-r}^{0}f_{Z}(z)
    dz+\frac{1}{r}\int_{0}^{r}f_{Z}(z) dz} =\left(\frac{\frac{1}{r}\int_{-r}^{0}f_{Z}(z)dz}{\frac{1}{r}\int_{0}^{r}f_{Z}(z) dz}+1\right)^{-1}\notag \\
    &=\left(\frac{f_{Z}^{-}(0)+\frac{1}{r}\int_{-r}^{0}(f_{Z}(z)-f_{Z}^{-}(0))dz}{f^{+}_{Z}(0)+\frac{1}{r}\int_{0}^{r}(f_{Z}(z)-f^{+}_{Z}(0)) dz}+1\right)^{-1}~.\label{eq:expansion-pir} 
\end{align}
Next, note that Assumption \ref{ass:A}(i') implies that $$\left|\frac{1}{r}\int_{0}^{r}(f_{Z}(z)-f_{Z}^{+}(0))dz\right|\le \frac{r}{2}C_P \quad \text{ and } \quad \left|\frac{1}{r}\int_{-}^{0}(f_{Z}(z)-f_{Z}^{-}(0))dz\right|\le \frac{r}{2}C_P ~.$$
Combining these two derivations we conclude that  
\begin{align*}
    \pi(r) &\le\left(\frac{f_{Z}^{-}(0)-\frac{r}{2}C_P}{f^{+}_{Z}(0)+\frac{r}{2}C_P}+1\right)^{-1} = \pi_f + \frac{\frac{r}{2}C_P}{f^{+}_Z(0)+f^{-}_Z(0)}~,\\
    \pi(r) &\ge\left(\frac{f_{Z}^{-}(0)+\frac{r}{2}C_P}{f^{+}_{Z}(0)\frac{r}{2}rC_P}+1\right)^{-1} = \pi_f - \frac{\frac{r}{2}C_P}{f^{+}_Z(0)+f^{-}_Z(0)}~.
\end{align*}
This implies the desired result.
\end{proof}

\section{Results under a mass point at the cut-off}\label{app:discontinuity}

In this section, we consider the asymptotic behavior of the proposed test when there is a mass point at the cut-off $\bar{z}=0$. As mentioned in Section \ref{sec:setup}, this mass point implies a violation of Assumption \ref{ass:A}, and so our formal results do not apply. On the other hand, a mass point at the cutoff is usually considered an extreme form of violation of the continuity of the density at the cut-off and, thus, should be regarded as part of $H_1$ in \eqref{eq:null}. The following result shows that the proposed test rejects with probability approaching one whenever there is a mass point zero.

\begin{theorem}\label{thm:allresults_mod}
Assume that $P\{Z=0\}>0$ and let $\alpha \in (0,1)$. If $q\geq q^*(\alpha)$ as in \eqref{eq:q-ast-alpha} and $\frac{q}{n} \to 0$,
  \begin{enumerate}[(a)]
      \item $S_{n}=q$ with probability approaching one.
      \item $\lim_{n\rightarrow \infty }E[\phi(S_n)]=1$.
  \end{enumerate}
\end{theorem}
\begin{proof}
Let $N_{n}\equiv \sum_{i=1}^{n}I\{ Z_{i}=0\} $. Note that $ N_{n}\geq q+1$ implies that $S_{n}=q$ so $T_{q}( S_{n}) = \sqrt{q}/2$. By this and $q\geq q^*(\alpha)$, $c_{q}( \alpha)<\sqrt{q}/2=T_{q}( S_{n}) $ and so $\phi ( S_{n}) =1$. Therefore, the desired results follow from showing that $ P\{ {\lim \inf}_{n\to\infty} \{ N_{n}\geq q+1\} \} =1$. To this end, note that $N_{n}\sim {\rm Bi}( n,P\{ Z=0\} ) $ so
\begin{align*}
    P\left\{ \underset{n\to\infty}{\lim \inf} \{ N_{n}\geq q+1\} \right\} =P\left\{  \underset{n\to\infty}{\lim \inf} \left\{ \frac{N_{n}}{n}\geq \frac{q+1}{n}\right\} \right\} \geq P\left\{  \underset{n\to\infty}{\lim \inf}\left\{ \frac{N_{n}}{n}\geq \frac{P\{ Z=0\}}{2} \right\} \right\} =1~,
\end{align*}
where the inequality holds for all $n$ large enough such that $ P\{ Z=0\}/2>(q+1)/n\to 0 $, and the last equality follows by the strong law of large numbers, i.e., $ N_{n}/n \overset{a.s.}{\to}P\{ Z=0\} >0$.
\end{proof}

It is relevant to note that Theorem \ref{thm:allresults_mod} applies to both asymptotic frameworks considered in the paper, i.e., it applies to fixed $q$ case as long as $q\geq q^*(\alpha)$, and it applies to large $q$ case provided that $q/n\to0$.

\section{Computational details on the data-dependent rule for \emph{q}}\label{app:q-irot}
\begin{figure}[th!]
\centering
  %!TEX root = ../RDDcontinuity.tex
\pgfplotsset{ %compat=1.8,
cycle list={
{draw=black, solid,very thick,blue},
{draw=black,dotted,very thick,orange}, 
{draw=black,densely dashed,very thick,blue},
{draw=black, dashdotted, very thick,green},
{only marks, mark=asterisk}}}

%\begin{subfigure}[t]{\textwidth}\centering  
    \begin{tikzpicture}[baseline]
         \begin{axis}[width=5in,height = 3.0in,xmin=6, xmax=150,ymin=0,ymax=6,ytick={1,2,3,4,5,6},xlabel=$q$,ylabel=Rejection Prob. in \%,legend style={at={(0.9,0.1)},anchor=south east}]
              \addplot  table[y = {rejection}]{qinformed-5.dat};
              \addlegendentry{$2\Psi_q(b_q(\alpha)-1)$}
              \addplot  table[y = {alpha}]{qinformed-5.dat};
              \addlegendentry{$\alpha$}
         \end{axis}
        %    \begin{customlegend}[legend columns=2,legend style={align=right,column sep=3ex},legend entries={{$\Psi_q(b_q(\alpha)$},{$\alpha$}}]
        %  \addlegendimage{solid,very thick,red}
        % \addlegendimage{densely dashed,very thick,blue} 
        % %\addlegendimage{dotted,very thick,orange}  
        % %\addlegendimage{dashdotted,very thick,green}
        % \end{customlegend}
    \end{tikzpicture}
%\end{subfigure}

% \begin{tikzpicture}
%         \begin{customlegend}[legend columns=2,legend style={align=center,column sep=3ex},legend entries={{$\Psi_q(b_q(\alpha)$},{$\alpha$}}]
%          \addlegendimage{solid,very thick,red}
%         \addlegendimage{densely dashed,very thick,blue} 
%         %\addlegendimage{dotted,very thick,orange}  
%         %\addlegendimage{dashdotted,very thick,green}
%         \end{customlegend}
% \end{tikzpicture}
    \caption{\footnotesize{The solid line is the limiting null rejection probability (in \%) of the non-randomized version of the test, $2\Psi_q(b_q(\alpha)-1)$, as a function of $q$. The dotted line is the nominal level of the test.}}
\label{fig:q-informed}
\end{figure}
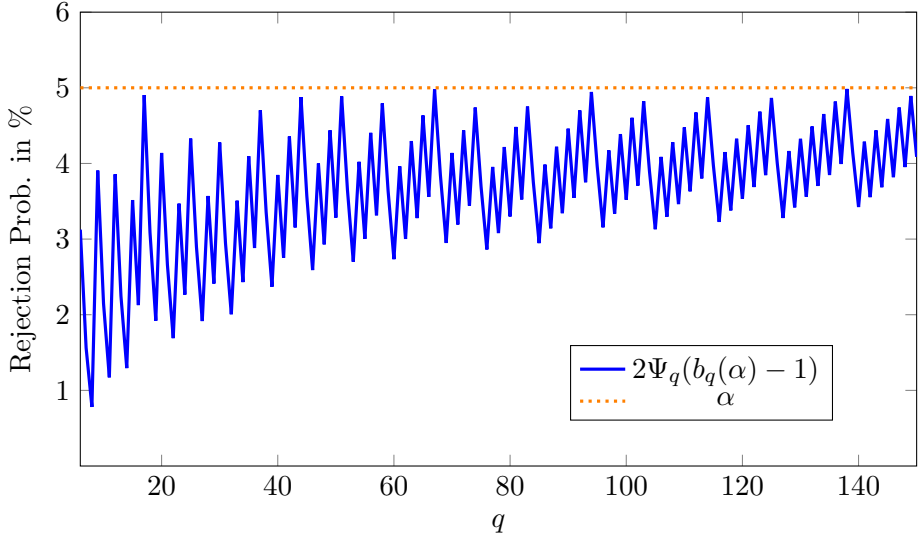

In the simulations of Section \ref{sec:simulations} and in the companion \verb+Stata+ package, the feasible informed rule of thumb is computed as follows. First, we compute   
\begin{equation*}\label{eq:qrot-feasible}
  \hat q_{\rm rot} = \left\lceil \max\left\lbrace q^*(\alpha),n^{1/2} \left( \hat \sigma \frac{4\phi^2_{\hat \mu,\hat \sigma}(\bar{z})}{\phi_{\hat \mu,\hat \sigma}(\hat \mu+\hat \sigma)}\right)^{2/3}\right\rbrace\right\rceil~,
\end{equation*}
where $q^*(\alpha)=1-\frac{\log \alpha}{\log 2}$, $\hat{\mu}$ is the sample mean of $\{Z_1,\dots,Z_n\}$, $\hat{\sigma}^2$ is the sample variance of $\{Z_1,\dots,Z_n\}$, $\bar{z}$ is the cut-off point, and $n$ is the sample size. In principle, the value $\hat q_{\rm rot}$ could be used to implement our test. However, this would ignore the non-monotonicity of the limiting null rejection probability of the non-randomized version of our test, which according to Theorem \ref{thm:main}, equals $2\Psi_q(b_q(\alpha)-1)$ with $b_q(\alpha)$ defined in \eqref{eq:bq}. Figure \ref{fig:q-informed} displays $2\Psi_q(b_q(\alpha)-1)$ for $\alpha=5\%$ as a function of $q$. The figure shows that $2\Psi_q(b_q(\alpha)-1)$ takes values very close to $\alpha$ for $q$ as low as 17 (i.e., $4.9\%$), but could be far from $\alpha$ for $q=19$ (i.e., $1.9\%$). We therefore propose an additional layer in the data-dependent way of choosing $q$ that guarantees that such a value delivers a local ``peak'' of $2\Psi_q(b_q(\alpha)-1)$ in Figure $\ref{fig:q-informed}$. 

To be concrete, we define $\hat q_{\rm irot}$ as 
 \begin{equation}\label{eq:q-irot-feasible}
    \hat q_{\rm irot}=\argmax_{q \in \mathcal N(\hat q_{\rm rot})} \Psi_q(b_q(\alpha)-1)~, 
  \end{equation}
where 
 \begin{equation}\label{eq:ENE}
\mathcal N(\hat q_{\rm rot})\equiv \{q\in \mathbf{N}: \max\{q^*(\alpha),\hat q_{\rm rot}-\lceil 4\log(\hat q_{\rm rot})\rceil \}\le q\le \hat q_{\rm rot} + \lceil 4\log(\hat q_{\rm rot})\rceil \}~.
  \end{equation}
The value of window size $\lceil 4\log(\hat q_{\rm rot})\rceil $ is the minimum number of points that are required to reach a local peak of $2\Psi_q(b_q(\alpha)-1)$ for values of $\alpha\in\{1\%,5\%,10\%\}$ and is such that, for large values of $\hat q_{\rm rot}$, the window gets larger to improve the chances of getting one of the peaks closer to $\alpha$ as $\hat q_{\rm rot}$ increases. A smaller window size may not guarantee one actually reaches a local peak. The value $\hat q_{\rm irot}$ defined in \eqref{eq:q-irot-feasible} is the one we use in the simulations of Section \ref{sec:simulations} and the default value in the companion \Verb+Rdcont+ \Verb+Stata+ package. 

\section{Surveyed papers on RDD}\label{app:survey}

Table \ref{tab:survey} displays the list of papers we surveyed in leading journals that use regression discontinuity designs. For a description on the criteria used to compile the list of papers in Table \ref{tab:survey}, see \citet[][Appendix E]{canay/kamat:18}.

\begin{table}[!th]
\begin{center}
\scalebox{0.45}{\hspace{-1cm}\begin{tabular}{lccccccccc} 
\hline \hline
Authors (Year) & & & Journal & & & (i): Mean Test & & & (ii): Density Test \\
\hline
Schmieder et al. (2016) & & & AER & & & \checkmark & & & \checkmark\\ 
Feldman et al. (2016) & & & AER & & & \checkmark & & & \checkmark\\  
Jayaraman et al. (2016) & & & AER & & & $\times$ & & & $\times$ 
\\Dell (2015) & & & AER & & & \checkmark & & & \checkmark\\ 
Hansen (2015) & & & AER & & & \checkmark & & & \checkmark\\ 
Anderson (2014) & & & AER & & & $\times$ & & & $\times$ 
\\Martin et al. (2014) & & & AER & & & $\times$ & & & $\times$ 
\\Dahl et al. (2014) & & & AER & & & \checkmark & & & \checkmark\\ 
Shigeoka (2014) & & & AER & & & \checkmark & & & $\times$ 
\\Crost et al. (2014) & & & AER & & & \checkmark & & & $\times$ 
\\Kostol and Mogstad. (2014) & & & AER & & & \checkmark & & & \checkmark\\ 
Clark and Royer (2013) & & & AER & & & \checkmark & & & $\times$ 
\\Brollo et al. (2013) & & & AER & & & \checkmark & & & \checkmark\\ 
Bharadwaj et al. (2013) & & & AER & & & \checkmark & & & \checkmark\\ 
Pop-Eleches and Urquiola (2013) & & & AER & & & \checkmark & & & \checkmark\\ 
Lacetera et al. (2012) & & & AER & & & $\times$ & & & \checkmark\\ 
Duflo et al. (2012) & & & AER & & & $\times$ & & & $\times$ 
\\Gopinath et al. (2011) & & & AER & & & \checkmark & & & \checkmark\\ 
Auffhammer and Kellogg (2011) & & & AER & & & $\times$ & & & $\times$ 
\\Duflo et al. (2011) & & & AER & & & $\times$ & & & $\times$ 
\\Ferraz and Finan (2011) & & & AER & & & $\times$ & & & $\times$ 
\\McCrary and Royer (2011) & & & AER & & & \checkmark & & & $\times$ 
\\Beland (2015) & & & AEJ:AppEcon & & & \checkmark & & & \checkmark\\ 
Buser (2015) & & & AEJ:AppEcon & & & \checkmark & & & \checkmark\\ 
Fack and Grenet (2015) & & & AEJ:AppEcon & & & \checkmark & & & \checkmark\\ 
Cohodes and Goodman (2014) & & & AEJ:AppEcon & & & \checkmark & & & \checkmark\\ 
Haggag and Paci (2014) & & & AEJ:AppEcon & & & \checkmark & & & \checkmark\\ 
Dobbie and Fryer (2014) & & & AEJ:AppEcon & & & \checkmark & & & \checkmark\\ 
Sekhri (2014) & & & AEJ:AppEcon & & & \checkmark & & & \checkmark\\ 
Schumann (2014) & & & AEJ:AppEcon & & & \checkmark & & & \checkmark\\ 
Lucas and Mbiti (2014) & & & AEJ:AppEcon & & & \checkmark & & & \checkmark\\ 
 \hline \hline
\end{tabular}%
\begin{tabular}{|lccccccccc} 
\hline \hline
Authors (Year) & & & Journal & & & (i): Mean Test & & & (ii): Density Test \\
\hline
Miller et al. (2013) & & & AEJ:AppEcon & & & \checkmark & & & \checkmark\\ 
Litschig and Morrison (2013) & & & AEJ:AppEcon & & & \checkmark & & & \checkmark\\ 
Dobbie and Skiba (2013) & & & AEJ:AppEcon & & & \checkmark & & & \checkmark\\ 
Kazianga et al. (2013) & & & AEJ:AppEcon & & & \checkmark & & & \checkmark\\ 
Magruder (2012) & & & AEJ:AppEcon & & & $\times$ & & & $\times$ 
\\Dustmann and Schnberg (2012) & & & AEJ:AppEcon & & & $\times$ & & & $\times$ 
\\Clots-Figueras (2012) & & & AEJ:AppEcon & & & \checkmark & & & \checkmark\\ 
Manacorda et al. (2011) & & & AEJ:AppEcon & & & \checkmark & & & \checkmark\\ 
Chetty et al. (2014) & & & QJE & & & \checkmark & & & \checkmark\\ 
Michalopoulos and Papaioannou (2014) & & & QJE & & & \checkmark & & & $\times$ 
\\Fredriksson et al. (2013) & & & QJE & & & \checkmark & & & \checkmark\\ 
Schmieder et al. (2012) & & & QJE & & & \checkmark & & & \checkmark\\ 
Lee and Mas (2012) & & & QJE & & & $\times$ & & & $\times$ 
\\Saez et al. (2012) & & & QJE & & & $\times$ & & & $\times$ 
\\Barreca et al. (2011) & & & QJE & & & $\times$ & & & $\times$ 
\\Almond et al. (2011) & & & QJE & & & \checkmark & & & \checkmark\\ 
Malamud and Pop-Eleches (2011) & & & QJE & & & \checkmark & & & \checkmark\\ 
Fulford (2015) & & & ReStat & & & \checkmark  & & & $\times$ 
\\Snider and Williams (2015) & & & ReStat & & & $\times$  & & & $\times$ 
\\Doleac and Sanders (2015) & & & ReStat & & & $\times$ & & & $\times$ 
\\Co\c sar et al. (2015) & & & ReStat & & & $\times$ & & & $\times$ 
\\Avery and Brevoort (2015) & & & ReStat & & & $\times$ & & & $\times$ 
\\Carpenter and Dobkin (2015) & & & ReStat & & & \checkmark & & & $\times$ 
\\Black et al. (2014) & & & ReStat & & & \checkmark & & & \checkmark\\ 
Anderson et al. (2014) & & & ReStat & & & $\times$ & & & $\times$ 
\\Alix-Garcia et al. (2013) & & & ReStat & & & \checkmark & & & $\times$ 
\\Albouy (2013) & & & ReStat & & & $\times$ & & & $\times$ 
\\Garibaldi et al. (2012) & & & ReStat & & & \checkmark & & & \checkmark\\ 
Manacorda (2012) & & & ReStat & & & \checkmark & & & \checkmark\\ 
Martorell and McFarlin (2011) & & & ReStat & & & \checkmark & & & \checkmark\\ 
Grosjean and Senik (2011) & & & ReStat & & & $\times$  & & & $\times$ \\
 \hline \hline
\end{tabular}%
}
\caption{{Survey of RDD empirical papers from 2011--2015 in the following journals: American Economic Review (AER), American Economic Journal: Applied Economics (AEJ:AppEcon), Quarterly Journal of Economics (QJE), and Review of Economics and Statistics (ReStat). Implications (i) and (ii) denote the testable implications proposed by \cite{lee:08} described in page \pageref{page:lee}. A checkmark indicates that the corresponding implication has been tested and a cross indicates otherwise.}}\label{tab:survey}
\end{center}
\end{table} 
%-------------------------------------------------------------

\newpage
\normalsize
\bibliography{RDD_Ref.bib}

\end{document}